\documentclass[usenatbib]{mn2e}
\usepackage{graphicx,times, amsmath, epsfig, color}
\usepackage[export]{adjustbox}

\newcommand{\sense}{\textsc{\small 21cmsense}}

\newcommand{\cmfastSM}{\textsc{\small 21CMFASTv2}}
\newcommand{\cmfast}{\textsc{\small 21CMFAST}}

\newcommand{\xz}{({\bf x}, z)}

\newcommand{\Ts}{T_{\rm S}}
\newcommand{\Tk}{T_{\rm K}}
\newcommand{\aveTs}{\bar{T}_{\rm S}}
\newcommand{\aveTk}{\bar{T}_{\rm K}}

\newcommand{\nf}{x_{\rm HI}}
\newcommand{\avenf}{\bar{x}_{\rm HI}}

\newcommand{\lya}{Ly$\alpha$}

\newcommand{\Msun}{M_\odot}
\newcommand{\Tvir}{T_{\rm vir}}

\newcommand{\Tcmb}{T_\gamma}
\newcommand{\avedelT}{\bar{\delta T_b}}
\newcommand{\delT}{\delta T_b}

\newcommand{\delNL}{\delta_{\rm nl}}
\newcommand{\Mmin}{M_{\rm min}}

\newcommand\lsim{\mathrel{\rlap{\lower4pt\hbox{\hskip1pt$\sim$}}
        \raise1pt\hbox{$<$}}}
\newcommand\gsim{\mathrel{\rlap{\lower4pt\hbox{\hskip1pt$\sim$}}
        \raise1pt\hbox{$>$}}}
\def\myputfigure#1#2#3#4#5%
{\vskip#5pt\makebox[0pt]{\hskip#2in
\includegraphics[width=#3\textwidth]{#1}}\vskip#4pt\hfill}

\newenvironment{packed_enum}{
\begin{enumerate}
  \setlength{\itemsep}{1pt}
  \setlength{\parskip}{0pt}
  \setlength{\parsep}{0pt}
}{\end{enumerate}}

\newcommand{\smallHII}{\textsc{\small Faint galaxies}}
\newcommand{\largeHII}{\textsc{\small Bright galaxies}}

\pdfoutput=1

\begin{document}

\title[EOS - Evolution Of 21cm Structure]{The Evolution Of 21cm Structure (EOS):\\
  public, large-scale simulations of Cosmic Dawn and Reionization}

\author[Mesinger et al.]{Andrei Mesinger\thanks{andrei.mesinger@sns.it}, Bradley Greig, \& Emanuele Sobacchi\\
Scuola Normale Superiore, Piazza dei Cavalieri 7, 56126 Pisa, PI, Italy
}

\voffset-.6in

\maketitle

\begin{abstract}
  We introduce the Evolution of 21-cm Structure (EOS) project: providing periodic, public releases of the latest cosmological 21-cm simulations.
  21-cm interferometry is set to revolutionize studies of the Cosmic Dawn (CD) and epoch of reionization (EoR), eventually resulting in 3D maps of the first billion years of our Universe.  Progress will depend on
  sophisticated data analysis pipelines, which are in turn tested on large-scale mock observations.
  %
  Here we present the 2016 EOS data release, consisting of the largest (1.6 Gpc on side with a 1024$^3$ grid), public 21-cm simulations of the CD and EoR.  We include calibrated, sub-grid prescriptions for inhomogeneous recombinations and photo-heating suppression of star formation in small mass galaxies.
  Leaving the efficiency of supernovae feedback as a free parameter, we present two simulation runs which approximately bracket the contribution from faint unseen galaxies.  Both models are calibrated to the Planck 2015 measurement of the Thompson scattering optical depth.
  From these two extremes, we predict that the duration of reionization (defined as a change in the mean neutral fraction from 0.9 to 0.1) should be between  $2.7 \lsim \Delta z_{\rm re} \lsim 5.7$.
  The large-scale 21-cm power during the advanced EoR stages can be different by up to a factor of $\sim10$, depending on the model.  This difference has a comparable contribution from: (i) the typical bias of sources; and (ii) a more efficient negative feedback in models with an extended EoR driven by faint galaxies.
  We also make detectability forecasts  with current and upcoming interferometers.
With a 1000h integration, HERA and SKA1-low should achieve a signal-to-noise of $\sim$few--hundreds throughout the EoR/CD, while in the maximally optimistic scenario of perfect foreground cleaning, all instruments should make a statistical detection of the cosmic signal.
  We also caution that our ability to clean foregrounds  determines the relative performance of narrow/deep vs. wide/shallow surveys expected with SKA1.
Our 21-cm power spectra, simulation outputs and visualizations are publicly available.

\end{abstract}


\section{Introduction}
\label{sec:intro}

Interferometry with the redshifted 21cm line is set to usher-in a ``Big Data'' revolution in astrophysical cosmology, eventually resulting in 3D maps of the $z\gsim6$ intergalactic medium (IGM).  Tracing the ionization and thermal state of cosmic gas, the 21cm signal is shaped by  radiation from the very first generations of astrophysical objects.  Thus the evolution of the spatial structure of the 21cm signal encodes information about the properties of early galaxies, black holes and the structure of the IGM.

First generation interferometers, such as the Low Frequency Array
(LOFAR; \citealt{vanHaarlem13})\footnote{http://www.lofar.org/}, the Murchison Wide 
 Field Array (MWA; \citealt{Tingay13})\footnote{http://www.mwatelescope.org/} and the 
 Precision Array for Probing the Epoch of Reionization 
 (PAPER; \citealt{Parsons10})\footnote{http://eor.berkeley.edu/}
  are currently taking data, hoping for a statistical detection of the epoch of reionization (EoR).  Second generation instruments, the Square Kilometre Array 
 (SKA)\footnote{https://www.skatelescope.org} and the Hydrogen Epoch of 
  Reionization Array (HERA)\footnote{http://reionization.org},
  will have the sensitivity to probe the 21cm signal over a much wider range of scales and redshifts.  These instruments are in the unique position to place powerful constraints on the first sources and sinks of radiation (e.g. \citealt{Pober14, Koopmans15, EW15}).  Advances in efficient 3D simulations of the cosmic signal will eventually allow us to constrain astrophysical cosmology from 21cm maps in an analogous fashion to standard Bayesian CMB analysis \citep{GM15}.

Unlocking the potential of 21cm interferometry hinges on our ability to dig out the cosmic signal from the data.  Thus in the near-term, sizeable efforts are being devoted to realistic data analysis, including antennae calibration, sky modelling, ionosphere, foreground and radio frequency interference (RFI) excision.  Testing analysis pipelines requires realistic mock observations in which the above sources of error are added to large-scale simulations of the cosmic signal.

In this work we introduce the Evolution Of 21cm Structure (EOS) project: providing periodic public releases of the latest, ``best-guess'' simulations of the cosmic 21cm signal.  These large-scale, high-resolution 21cm maps can aid data pipeline construction, forecasting, observational strategies, instrument design, as well as calibrating more approximate, fast simulations.  The maps and associated visualizations are available for download at http://homepage.sns.it/mesinger/EOS.html.


Here we provide two simulation outputs which crudely bracket uncertainties in source models.  We caution however that this is by no means representative of all current uncertainties, and the EOS outputs should only be viewed as a compliment to exhaustive astrophysical parameter exploration with faster codes \citep{GM15}.
Following improvements in both theoretical modelling and $z\gsim6$ observations, we anticipate periodic releases on approximately a yearly time-scale.

The structure of this paper is as follows.  In \S \ref{sec:sim}, we describe the simulations and their set-up. In \S \ref{sec:IGM}, we show the evolution of IGM properties in our models. In \S \ref{sec:results}, we present the main results: 21-cm predictions from our models, including estimates of their detectability with current and upcoming interferometers.
We conclude in \S \ref{sec:conc}.  Unless stated otherwise, we quote all quantities in comoving units. We adopt the background cosmological parameters: ($\Omega_\Lambda$, $\Omega_{\rm M}$, $\Omega_b$, $n$, $\sigma_8$, $H_0$) = (0.69, 0.31, 0.048, 0.97, 0.82, 68 km s$^{-1}$ Mpc$^{-1}$), consistent with recent results from the Planck mission \citep{Planck15}.

\section{Simulations}
\label{sec:sim}

The 21-cm signal is usually represented in terms of the offset of the 21cm brightness temperature from the CMB temperature, $\Tcmb$, along a line of sight (LOS) at an observed frequency $\nu$ (c.f. \citealt{FOB06}):
\begin{align}
\label{eq:delT}
\nonumber \delT(\nu) = &\frac{\Ts - \Tcmb}{1+z} (1 - e^{-\tau_{21}}) \approx \\
\nonumber &27 \nf (1+\delNL) \left(\frac{H}{dv_r/dr + H}\right) \left(1 - \frac{\Tcmb}{\Ts} \right) \\
&\times \left( \frac{1+z}{10} \frac{0.15}{\Omega_{\rm M} h^2}\right)^{1/2} \left( \frac{\Omega_b h^2}{0.023} \right) {\rm mK},
\end{align}
\noindent where $T_S$ is the gas spin temperature, $\nf$ is the neutral fraction, $\tau_{21}$ is the optical depth at the 21-cm frequency $\nu_{21}$, $\delNL({\bf x}, z) \equiv \rho/\bar{\rho} - 1$ is the evolved (Eulerian) density contrast, $H(z)$ is the Hubble parameter, $dv_r/dr$ is the comoving gradient of the line of sight component of the comoving velocity, and all quantities are evaluated at redshift $z=\nu_{21}/\nu - 1$.

To simulate the 21-cm signal, we use a modified version of the publicly available code \cmfast\footnote{http://homepage.sns.it/mesinger/Sim.html} \citep{MF07, MFC11}.  Specifically, we use \cmfastSM\ \citep{SM14},  which incorporates calibrated, sub-grid prescriptions for inhomogeneous recombinations and photo-heating suppression of the gas fraction in small halos.  The simulation boxes are 1.6 Gpc on a side, with a resolution of 1024$^3$.  The cosmological terms in eq. (\ref{eq:delT}), namely the density field and its corresponding redshift space distortions, are computed using perturbation theory (e.g. \citealt{Zeldovich70}).  The $\nf$ and $\Ts$ terms depend on both cosmology and radiation fields. The neutral fraction depends on ionizing UV and X-ray radiation, as well as inhomogeneous recombinations.  The spin temperature interpolates between the gas temperature, $\Tk$, which is set by X-ray heating (e.g. \citealt{MO12}), and the CMB temperature, $\Tcmb$.  The strength of this coupling during the Cosmic Dawn is set by the Lyman alpha background [so-called Wouthuysen-Field (WF) coupling; \citealt{Wouthuysen52, Field58}].

Therefore, to model the 21cm-signal during the EoR and Cosmic Dawn, we need three radiation fields: (i) ionizing UV; (ii) X-ray; (iii) \lya.  Below we summarize how these are computed in turn (for more details readers are encouraged to see \citealt{MF07, MFC11, MFS13, SM14}).  As already mentioned, a full astrophysical parameter space exploration is beyond the scope of this work; instead {\it we present two runs, approximately bracketing the contribution from  faint, unseen galaxies.}

\subsection{UV Ionization}
\label{sec:EoR}

\cmfast\ computes reionization through an excursion-set approach \citep{FZH04}, applied to a non-linear 3D density field.  The excursion set approach classifies regions as either neutral or ionized, based on whether their cumulative output of ionizing photons is greater than the number of baryons plus cumulative recombinations (see \S \ref{sec:xray} for our treatment of X-ray ionization). Specifically, a cell located at spatial position and redshift, ($\textbf{x}$, $z$), is flagged as ionized if:
\begin{equation}
\label{eq:ion_crit_coll}
\zeta f_{\rm coll}(\textbf{x}, z, R, \bar{M}_{\rm min})\geq 1+\bar{n}_{\rm rec}(\textbf{x}, z, R)
\end{equation}
where $f_{\rm coll}\left(\textbf{x},z, R, \bar{M}_{\rm min}\right)$ is the fraction of collapsed matter (computed according to \citealt{BL04} and \citealt{MFC11}) inside a sphere of radius $R$ residing in halos larger than $\bar{M}_{\rm min}$ (see \S \ref{sec:sources}), $\bar{n}_{\rm rec}$ is the average cumulative number of recombinations (see \S \ref{sec:sinks}), and $\zeta$ is an ionizing efficiency:
\begin{equation}
\label{eq:zeta}
\zeta = 20 \bigg(\frac{N_\gamma}{4000}\bigg) \bigg(\frac{f_{\rm esc}}{0.1}\bigg) \bigg(\frac{f_\ast}{0.05}\bigg) \bigg(\frac{f_{\rm b}}{1}\bigg)~ ,
\end{equation}
where $N_\gamma$ is the number of ionizing photons per stellar baryon, $f_{\rm esc}$ is the fraction of UV ionizing photons that escape into the IGM, $f_\ast$ is the fraction of galactic gas in stars, and $f_{\rm b}$ is the fraction of baryons inside the galaxy with respect to the cosmic mean $\Omega_{\rm b}/\Omega_{\rm m}$ (note that some works include the total number of homogeneous recombinations inside the definition of $\zeta$).

The LHS of eq. (\ref{eq:ion_crit_coll}) corresponds to the time-integrated number of photons per baryon inside a given region, while the RHS corresponds to the time-integrated number of recombinations plus one.  Both of these quantities depend on the local properties of the cosmic gas, and are tracked in each cell of our simulation.  We discuss how these are computed in greater detail below.

\subsubsection{Ionization sources}
\label{sec:sources}

Reionization sources are taken to be galaxies hosted by halos above some spatially-dependent minimum halo mass, $M_{\rm min}$.  The value of this threshold is set by the ability of the cosmic gas to efficiently cool and form stars inside the halo, depending on cooling, photo-heating and supernovae (SNe) feedback. Specifically, we take:
\begin{equation}
  \label{eq:Mmin_components}
  \Mmin (\textbf{x}, z) = \max \left[ M_{\rm cool}, M_{\rm photo}, M_{\rm SNe} \right] ~.
\end{equation}
Here $M_{\rm cool}$ corresponds to the virial temperature, taken to be $T_{\rm vir}\gsim 2\times10^{4}\text{ K}$, required for efficient cooling through atomic hydrogen\footnote{Molecular hydrogen cooling is expected to be strongly suppressed well before the bulk of reionization (e.g. \citealt{HRL97, HF12, Fialkov13}).}. Note that the virial temperature can be related to the halo mass through (e.g. \citealt{BL01}):
\begin{align}
M &= 10^{8} h^{-1} \left(\frac{\mu}{0.6}\right)^{-3/2}\left(\frac{\Omega_{\rm m}}{\Omega^{z}_{\rm m}}
\frac{\Delta_{\rm c}}{18\pi^{2}}\right)^{-1/2} \nonumber \\
& \times \left(\frac{T_{\rm vir}}{1.98\times10^{4}~{\rm K}}\right)^{3/2}\left(\frac{1+z}{10}\right)^{-3/2}M_{\sun},
\end{align}
where $\mu$ is the mean molecular weight, $\Omega^{z}_{\rm m} = \Omega_{\rm m}(1+z)^{3}/[\Omega_{\rm m}(1+z)^{3} + 
\Omega_{\Lambda}]$, and $\Delta_{c} = 18\pi^{2} + 82d - 39d^{2}$ where $d = \Omega^{z}_{\rm m}-1$.

In reionized regions, the gas reservoir to form stars can be depleted with respect to the cosmic mean ($f_b < 1$), due to heating from the UV background (UVB; e.g. \citealt{SGB94, HG97, MD08, OGT08, PS09}).  Halos with masses above $M_{\rm photo}$ retain enough gas (defined so that $f_b \geq 1/2$; c.f eq. \ref{eq:zeta}) to continue efficiently forming stars inside HII regions.  We use the following empirical formula, whose form is motivated by linear theory \citep{SM13a}:
\begin{align}
\label{eq:M_crit}
M_{\rm photo}(\textbf{x}, z) = &M_{0} \left(\frac{\Gamma_{\rm halo}}{10^{-12}\text{ s$^{-1}$}}\right)^a
\left(\frac{1+z}{10}\right)^{b}\left[1-\left(\frac{1+z}{1+z_{\rm re}}\right)^{c}\right]^{d}
\end{align}
where $\left(M_{0}, a, b, c, d\right) = \left(3.0\times 10^{9}M_{\odot}, 0.17, -2.1, 2.0, 2.5\right)$ are fitted to suites of 1D collapse simulations exploring a wide parameter space.  At the high redshifts of interest, this result was found to be consistent with a more sparsely sampled parameter space of 3D hydrodynamic simulations \citep{NM14}.
The local value of $M_{\rm photo}(\textbf{x}, z)$ depends on the cells' reionization redshift $z_{\rm re}$, and also (weakly) on the impinging photo-ionization rate, $\Gamma_{\rm halo}$ (e.g. \citealt{SM14}):
\begin{align}
\label{eq:self_cons_J}
\nonumber \Gamma_{\rm halo}(\textbf{x}, z) &= f_{\rm bias}\times \Gamma_{\rm HII} \\
&\approx f_{\rm bias} \left(1+z\right)^2 \lambda_{\rm mfp} \sigma_{\rm H}\frac{\alpha}{\alpha+\beta}\bar{n}_{\rm b} \zeta \frac{df_{\rm coll}}{dt} ~ .
\end{align}
where $\Gamma_{\rm HII}$ is the average photo-ionization rate in the surrounding HII region, $f_{\rm bias} \sim 2$ is the mean enhancement of the UVB at galaxy locations due to the clustering of sources \citep{MD08}, $\lambda_{\rm mfp}$ is the mean free path of the local HII region (set by either the HII morphology or the instantaneous, inhomogeneous recombinations discussed in the next section), $\bar{n}_{\rm b}$ is the mean baryon number density inside the HII region, and we assume a photo-ionization cross-section $\sigma\left(\nu\right)=\sigma_{\rm H}\left(\nu/\nu_{\rm H}\right)^{-\beta}$ with $\sigma_{\rm H}=6.3\times 10^{-18}\text{ cm$^2$}$ and $\beta =2.75$.  The above also assumes that the ionizing emissivity is spectrally distributed as $\nu^{-\alpha}$ (we use $\alpha=5$, corresponding to a stellar-driven UV spectrum; e.g. \citealt{TW96}).
Thus each simulation cell keeps track of its own value of $M_{\rm photo}$, which depends on its photo-ionization history.

\begin{figure*}
\vspace{-1\baselineskip}
{
  \includegraphics[width=0.45\textwidth]{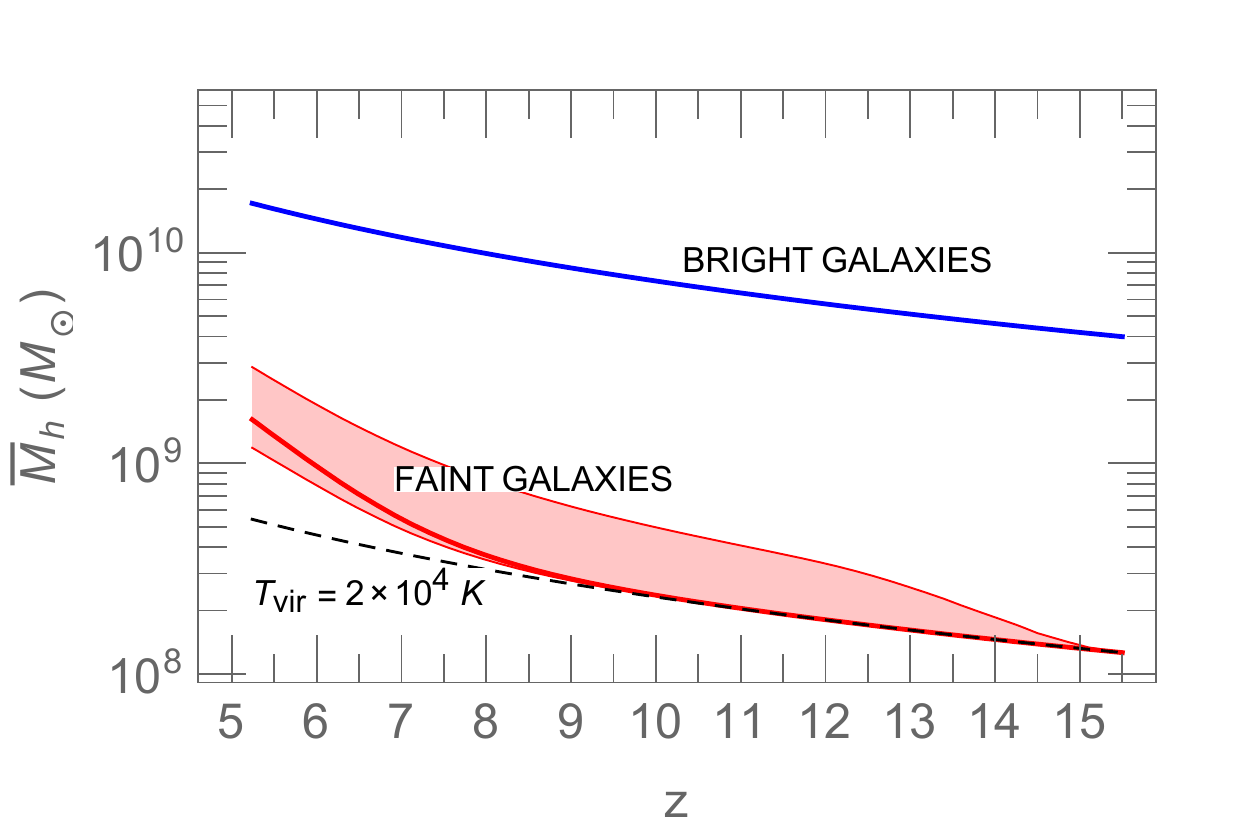}
  \includegraphics[width=0.45\textwidth]{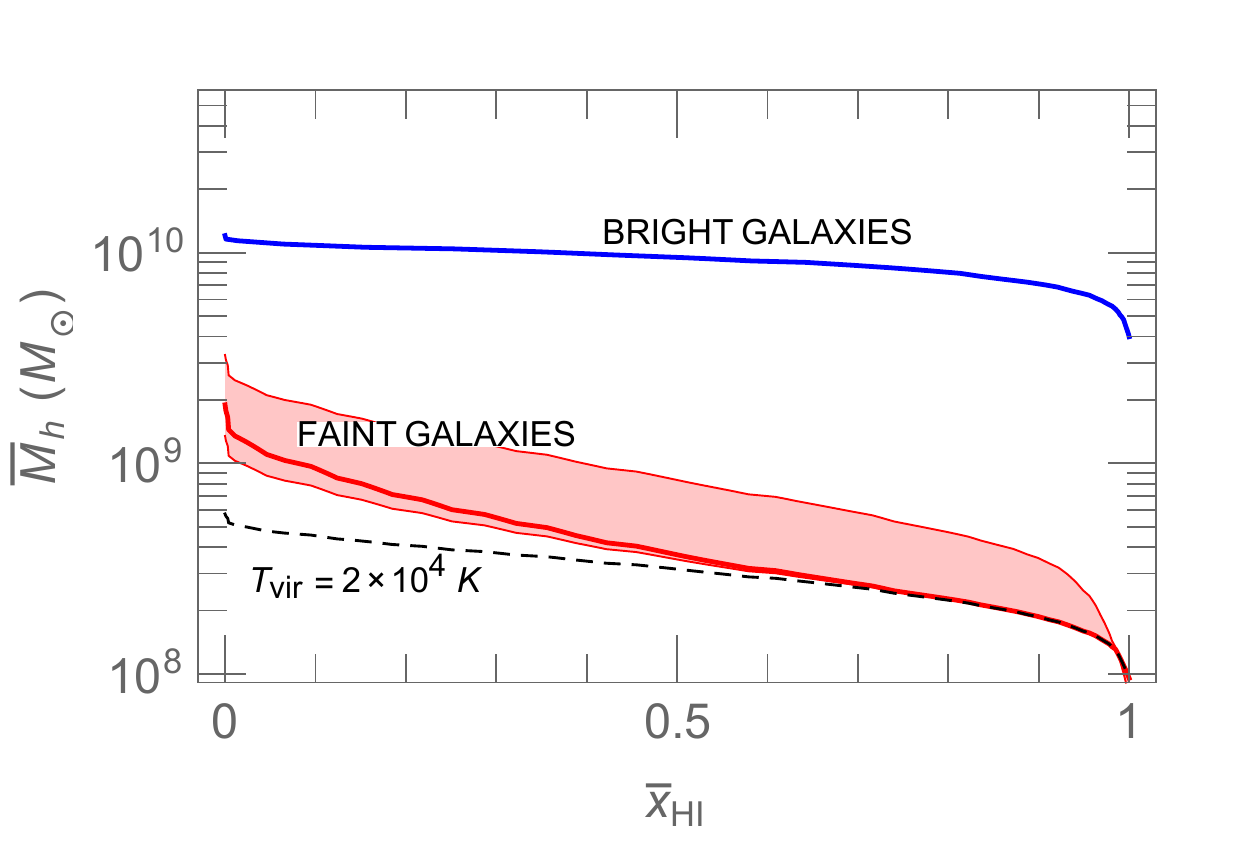}
}
\caption{
  The average mass of halos which host the dominant ionizing sources.  Specifically, we plot the mass (i.e. luminosity) averaged halo mass in our models.  In the left panel we show its evolution vs redshift, while in the right panel we show its evolution vs $\avenf$.  The red shaded region corresponds to the 1$\sigma$ spatial scatter at a given redshift, which in our model is driven by the inhomogeneity of photo-heating suppression of the galaxy gas content.  The dashed black curve corresponds to a constant virial temperature of $2\times10^4$ K; by construction, the \largeHII\ model follows the corresponding curve for $2\times10^5$ K.
}
\label{fig:Mhalo}
\vspace{-0.5\baselineskip}
\end{figure*}

The final term in eq. (\ref{eq:Mmin_components}),  $M_{\rm SNe}$, corresponds to SNe feedback in high-$z$ galaxies (e.g. \citealt{SH03}).
How SNe deposit energy in their surroundings and the impact this has on future star formation is one of the largest outstanding questions in early Universe astrophysics.  Indeed, we hope that the sensitivity of upcoming cosmological 21-cm measurements to $M_{\rm min}$ (e.g. \citealt{McQuinn07, GM15, Geil15, Dixon15}) will indirectly help us understand this process, which is intractable from first principles.  In this work, we take $M_{\rm SNe}$ to be a free parameter, adopting two extreme values which bracket the allowed range, and then {\it adjust the ionizing efficiency $\zeta$ to be consistent with the measurement of the Thompson scattering optical depth to the CMB, $\tau_e = 0.066 \pm 0.16$ \citep{Planck15}.}
\begin{itemize}
\item {\bf \smallHII} -- This simulation run corresponds to the limit of inefficient or halo-mass independent SNe feedback, i.e. $M_{\rm SNe} < \max[M_{\rm cool}, M_{\rm photo}]$.  Star forming galaxies in this model are hosted predominantly by fairly small halos close to the atomic cooling threshold of $\Tvir \sim 10^4$ K (c.f. Fig. \ref{fig:Mhalo}).  The dominant galaxies therefore are relatively abundant, have a small bias, form early and evolve slowly.  For this model, we set the ionizing efficiency to be $\zeta=20$, resulting in $\tau_e = 0.069$.\\
\item {\bf \largeHII} --  This simulation run corresponds to extremely efficient SNe feedback,  sharply suppressing the star formation inside halos with $\Tvir < 2 \times 10^5$ K.  This roughly translates to the limit of currently observed high-$z$ Lyman break galaxy (LBG) candidates, obtained through abundance matching (e.g. \citealt{KF-G12, GM15, Atek15, SF15}), assuming a duty cycle of unity (e.g. \citealt{Barone-Nugent14}).
  For this run, we take the ionizing effciency to be $\zeta=200$, resulting in $\tau_e=0.066$. 
\end{itemize}

Although both runs can be considered as extreme source models, we expect that {\it the predictions of the \smallHII\ model are closer to the true signal}.  Firstly, we note that the choice of $\zeta=20$ in the \smallHII\ run is consistent with the fiducial values shown in eq. (\ref{eq:zeta}).  In contrast, the extreme value of $\zeta=200$ used in the \largeHII\ run would imply an ionizing escape fraction close to unity for the relatively bright LBGs, which is inconsistent with observations at $z\sim3$ (e.g. \citealt{KF-G12, Cooke14, Khaire15}). Secondly, although SNe likely do play an important role in regulating star formation at high redshifts, it is not clear whether this results in a sharp suppression of star formation at a given halo mass (or $\Tvir$) scale.  Indeed our formalism in eq. (\ref{eq:Mmin_components}) assumes a sharp (step function) suppression of star formation below the given threshold.  A sharp transition is a justified assumption for the cooling threshold (e.g. \citealt{BL01}), and the threshold for photo-heating suppression (e.g. \citealt{SM13a, NM14}).  However this assumption is uncertain for the case of SNe feedback.  Note that {\it halo mass independent scalings of the star formation are absorbed in the $\zeta$ parameter}. Thus SNe feedback whose suppression of star formation is roughly independent of halo mass would still be described by our \smallHII\ galaxy model, with a correspondingly low value of $f_\ast$.  Lastly, even if star formation drops rapidly towards small mass halos due to SNe feedback, the shear increase in their abundance can compensate for this.  
For example, the calibrated semi-analytic models (SAMs) of galaxy formation during the EoR by \citet{Mutch15} still result in reionization being dominated by $M\sim10^{9-10} \Msun$ halos, despite the fact that the average value of $f_\ast$ is over a factor of ten lower inside these objects.

In order to explicitly quantify the dominant galaxy populations in our two source models, in Fig. \ref{fig:Mhalo}, we plot the evolution of the mass-weighted (which for our model is analogous to weighing by the ionizing luminosity) average halo mass contributing to reionization as a function of redshift ({\it left panel}) and $\avenf$ ({\it right panel}).  The red shaded region corresponds to the 1$\sigma$ spatial scatter at a given redshift, which in our model is driven by the inhomogeneity of $M_{\rm photo}(\textbf{x}, z)$.  The dashed black curve corresponds to a constant virial temperature of $2\times10^4$ K; by construction, the \largeHII\ model follows the corresponding curve for $2\times10^5$ K.

From the figure, we see explicitly that $M_{\rm SNe}(z) \gg M_{\rm photo}(\textbf{x}, z)$ in the \largeHII\ model; thus there is virtually no spatial scatter in $M_{\rm min}$.  Moreover, reionization occurs so quickly in this model that one can accurately approximate the EoR as being driven by galaxies residing in halos with masses of $10^{10} \Msun$.

On the other hand, the \smallHII\ model shows a factor of 2--3 spatial scatter in the typical halo mass of reionizing galaxies, driven by the interplay of recombinations and photo-heating feedback on gas accretion.  As time progresses, less gas is available to form stars in galaxies close to the atomic cooling threshold and gas clumps which were the first to be reionized start recombining, resulting in a rapid rise in the typical halo which hosts galaxies at $z\lsim7$.  As we shall see below, this feedback damps the otherwise expected rapid rise in the ionizing background at these redshifts \citep{SM14}.

\subsubsection{Ionization sinks}
\label{sec:sinks}

The RHS of eq. (\ref{eq:ion_crit_coll}) contains the cumulative number of recombinations, averaged over a given scale $R$.  We compute this term following \citet{SM14}:
\begin{equation}
\label{eq:n_rec}
\bar{n}_{\rm rec}(\textbf{x}, z)=\left\langle\int_{z_{\rm re}}^{z}\frac{dn_{\rm rec}}{dt}\frac{dt}{dz}dz\right\rangle_{\rm R} ~ ,
\end{equation}
with each ionized cell keeping track of its recombination rate per baryon:
\begin{equation}
\label{eq:dn_rec}
\frac{d n_{\rm rec}}{dt}(\textbf{x}, z)=\int_0^{\infty} P_{\rm V} \alpha_{\rm B} (1-x_{\rm HI})^{2} \bar{n}_{\rm H} \Delta d\Delta ~.
\end{equation}
Here, $P_{\rm V}(\Delta ~| ~z, \Delta_{\rm cell})$ is the sub-grid, volume weighted distribution of non-linear overdensities $\Delta\equiv n(\textbf{x})/\bar{n}$ [we take the fits presented in \citet{MHR00}, adjusted for the cell's average density $\Delta_{\rm cell}$ according to \citet{SM14}], $\alpha_{\rm B}$ is the case-B recombination coefficient at a temperature of 10$^4\text{ K}$ \citep{Osterbrock89}, and the neutral fraction corresponding to a given sub-grid density, $x_{\rm HI}(\Delta ~| ~z, \Gamma_{\rm HII})$ is computed assuming ionization equilibrium and the empirical self-shielding prescription of \citet{Rahmati13}.  We note that our sink prescription results in a residual HI mass fraction of $\sim$ 4\% (1\%) post EoR for the \smallHII\ (\largeHII) model.

\subsection{X-ray radiation}
\label{sec:xray}

Although X-rays can produce some level of pre-reionization (e.g. \citealt{RO04, Dijkstra12, McQuinn12, MFS13}), their dominant role is in heating the IGM.  Following an X-ray photo-ionization of H or He, the primary electron's energy gets used initially to create secondary ionizations.  However, at ionization fractions greater than a few percent, the bulk of the primary's energy gets converted to heat via free-free interactions (e.g. \citealt{SvS85, FS10, VEF11}).  Empirical relations of the X-ray luminosity to star-formation rate from local galaxies suggest that X-rays heated the IGM to temperatures above the CMB before reionization (e.g. \citealt{MO12, Pacucci14}), thus driving the 21-cm line from being seen in absorption to being seen in emission against the CMB.\footnote{Our simulations include X-ray heating, adiabatic cooling/heating, and Compton heating/cooling.  We do not account for shock heating, which convergence tests with hydrodynamic simulations have shown to be subdominant at $z\gsim10$ (e.g. Fig. A1 in \citealt{MO12}).}

As described in \citet{MFC11, MFS13}, we follow the ionization and heating evolution equations for each simulation cell, taking from \citet{FS10} the estimate of the fraction of the primary electron's energy which goes into (i) ionizations, (ii) heating and (iii) \lya\ emission.  The heating and ionization evolution is computed from the angle-averaged specific intensity, $J(\textbf{x}, \nu, z)$, (in erg s$^{-1}$ Hz$^{-1}$ pcm$^{-2}$ sr$^{-1}$; a prefix of 'p' in 'pcm' denotes proper units), by integrating the comoving specific emissivity (energy per unit time per unit frequency per unit comoving volume), $\epsilon_{h \nu}(\textbf{x}, \nu_e, z')$ back along the light-cone:
\begin{equation}
\label{eq:J}
J(\textbf{x}, \nu, z) = \frac{(1+z)^3}{4\pi} \int_{z}^{\infty} dz' \frac{c dt}{dz'} \epsilon_{h \nu}  e^{-\tau} ~ ,
\end{equation}
where $e^{-\tau}$ is the probability a photon emitted at $z'$ survives to reach $z$\footnote{The IGM optical depth (dominated by photo-ionizations of H and He) between $z$ and $z'$ is computed accounting for both UV and X-ray ionizations.  Motivated by the long mean free path of typical X-ray photons, we also assume a spatially homogeneous value for $\tau$.},  
and the comoving specific emissivity
is evaluated in the emitted frame, $\nu_e = \nu (1+z')/(1+z)$.  Assuming that the source luminosities follow a power law with spectral index $\alpha_{\rm X}$ beyond some threshold $\nu_0$ (corresponding to absorption within the galaxy), and adopting a normalization of $N_{\rm X}$ X-ray photons produced per stellar baryon, we can write:
\begin{equation}
\label{eq:emissivity}
\epsilon_{h \nu}(\textbf{x}, \nu_e, z') = \frac{\alpha_{\rm X} h}{\mu m_p} N_{\rm X} \left( \frac{\nu_e}{\nu_0} \right)^{-\alpha_{\rm x}} f_{\rm esc} \left[ f_{\ast} \rho_{\rm crit, 0} \Omega_b \frac{d f_{\rm coll}(z')}{dt} \right] ~ ,
\end{equation}
$\mu m_p$ is the mean baryon mass, $\rho_{\rm crit, 0}$ is the current critical density, $f_\ast$ is fraction of baryons converted into stars.  The quantity in the square brackets is the star formation rate density at $z'$ (mass of stars formed per unit time per unit comoving volume).
Here we take $\nu_0 = 0.3$ keV, $\alpha_{\rm X}=1.5$, and $N_{\rm X}=0.2$.  These values for X-ray sources are motivated by observations of local star-forming galaxies, in which
high mass X-ray binaries and the hot IGM contribute equally to the soft X-ray luminosity relevant for heating the IGM (see \citealt{Pacucci14} and references therein).

\subsection{\lya\ radiation}
\label{sec:lya}

As discussed above, the Lyman alpha background is responsible for coupling the spin temperature and the kinetic temperature during the Cosmic Dawn.
It has two contributors:  (i)  X-ray excitation of HI, which scales with the X-ray intensity, eq. (\ref{eq:J}); and (ii) stellar emission of photons in the Lyman bands.
For reasonable choices of $N_{\rm X} \sim 1000$, (ii) dominates the \lya\ background (e.g. \citealt{MFS13}).  

Because of the high resonant optical depth of neutral hydrogen, photons redshifting into any Lyman-$n$ resonance at $\xz$ will be absorbed in the IGM.  They then quickly and locally cascade with a fraction $f_{\rm recycle}(n)$ passing through \lya\ and inducing strong coupling \citep{Hirata06, PF06}.  Therefore, the direct stellar emission component of the \lya\ background can be estimated with a sum over the Lyman resonance backgrounds (e.g. \citealt{BL05_WF}), again integrating back along the light cone as in eq. (\ref{eq:J}).  We assume a Population II stellar emission (e.g. \citealt{BL05_WF}, and $f_\ast=0.05$ as in eq. (\ref{eq:zeta}).  For further details on the calculation, please see \citet{MFC11}.
We note that the relevant mean free paths of photons just blueward of Ly$\beta$ are of order 100 Mpc, making WF coupling the most homogeneous of the radiation-dominated epochs (e.g. \citealt{PF07}).

\section{IGM properties}
\label{sec:IGM}

\begin{figure}
\vspace{-1\baselineskip}
{
  \includegraphics[width=0.45\textwidth]{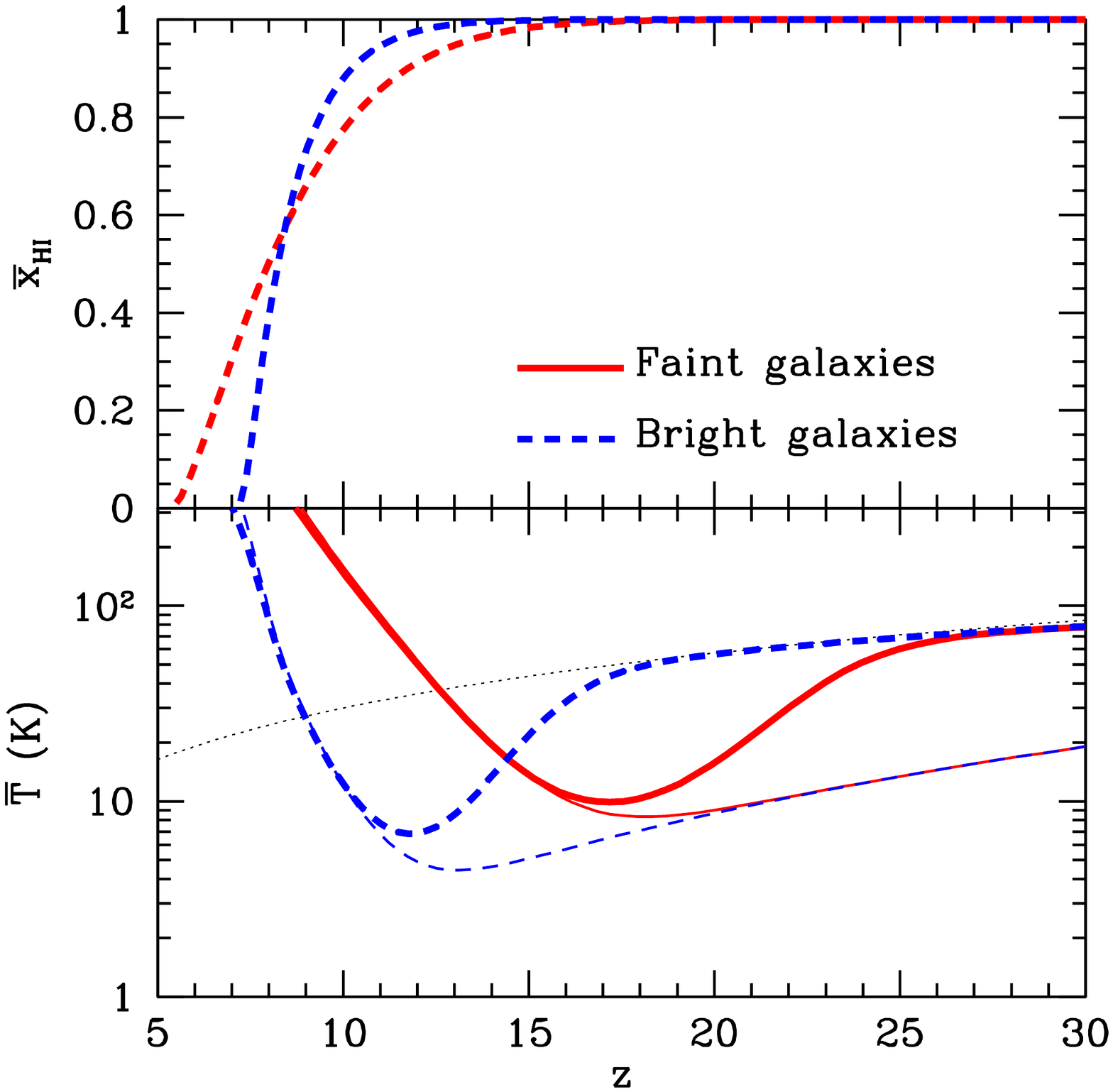}
}
\caption{
Redshift evolution of the volume averaged neutral fraction ({\it top}) and temperature ({\it bottom}).  The thick (thin) curves in the bottom panel correspond to $\aveTs$ ($\aveTk$), with the black dotted line showing $T_\gamma$. Red curves correspond to the \smallHII\ run, while blue curves correspond to the \largeHII\ run.  As discussed in the text, the ionizing efficiency of the galaxies is adjusted to obtain comparable values of $\tau_e = $0.069 (0.066) for the \smallHII\ (\largeHII) models.  However, the evolution of IGM properties is much more rapid in the \largeHII\ run, as it is driven by galaxies hosted by rare, rapidly-forming halos.
}
\label{fig:global}
\vspace{-0.5\baselineskip}
\end{figure}

In Fig. \ref{fig:global} we present the redshift evolution of the volume averaged neutral fraction ({\it top}) and temperature ({\it bottom}).  The thick (thin) curves in the bottom panel correspond to $\aveTs$ ($\aveTk$), with the black dotted line showing $T_\gamma$.  The most noticeable trend in Fig. \ref{fig:global} is that IGM properties evolve more rapidly when the radiation fields are driven by galaxies inside more rare, rapidly-forming halos. Although the \smallHII\ and \largeHII\ models have comparable midpoints of reionization (by construction), their durations are very different.  Specifically the transition from 90\% neutral to 10\% neutral occurs over a redshift interval of $\Delta_{0.9-0.1} z_{\rm re} \approx$ 5.7 in the \smallHII\ model to $\Delta_{0.9-0.1} z_{\rm re} \approx$ 2.7 in the \largeHII\ model.  Given that these models are extremes, this range is a fairly robust prediction for the duration of reionization.

The differences in how fast IGM properties evolve is also noticeable in the average temperatures, shown in the bottom panel.  We note that although the ionizing photon production efficiency, $\zeta$, was adjusted to result in the same midpoint of reionization in the two models, no such tuning was performed for the temperature evolution.  As a result, the same X-ray astrophysical parameters ($\nu_0$, $N_{\rm X}$, $\alpha_{\rm X}$) result in later heating when galaxies are hosted in more massive, late-appearing halos.  For example,  in the \largeHII\ model the spin temperature surpasses the CMB temperature at a time when reionization is well underway, with $\avenf\sim0.6$.  In contrast, the \smallHII\ model has this transition occurring much earlier, when $\avenf\sim0.9$.  As we shall see below, this difference in the overlap of EoR and epoch of X-ray heating (EoX) has important consequences for the detectability of the 21-cm signal.

\begin{figure}
\vspace{-1\baselineskip}
{
  \includegraphics[width=0.45\textwidth]{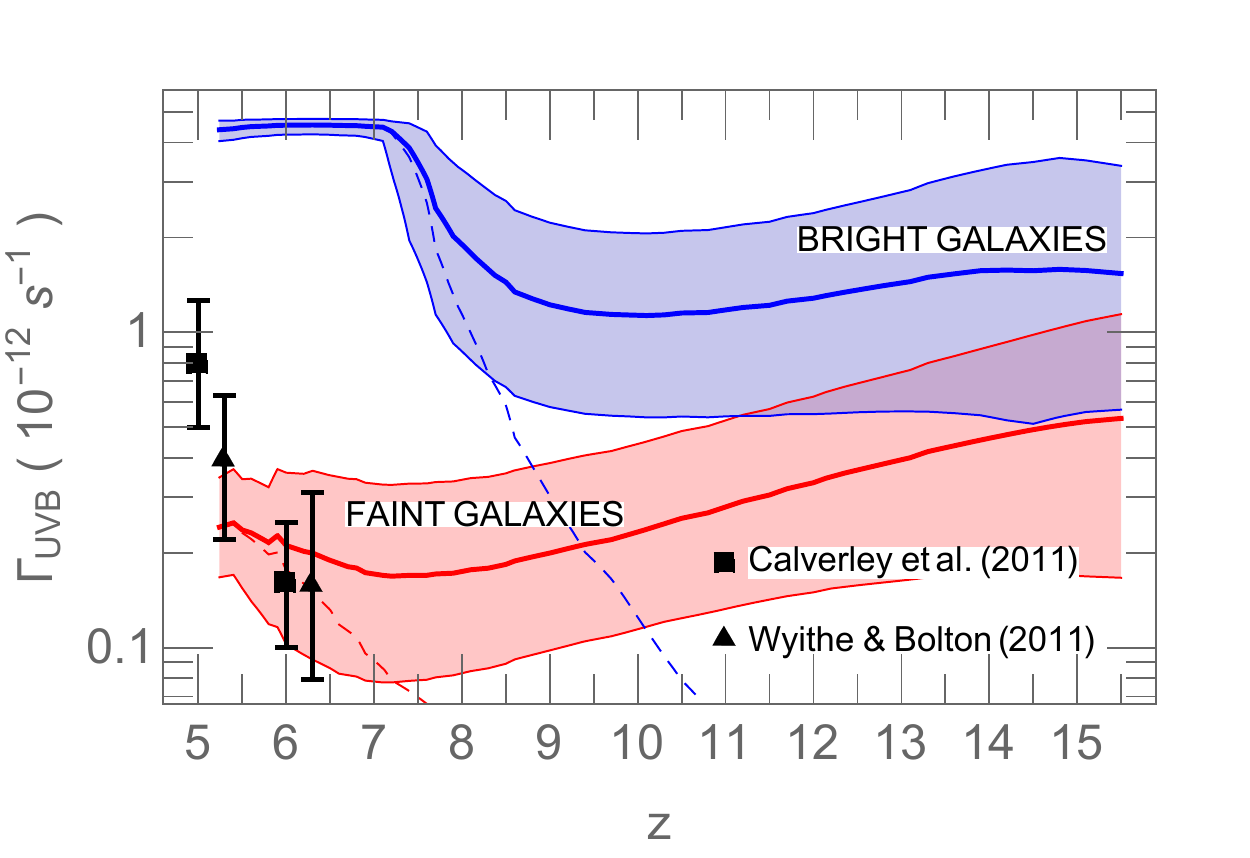}
}
\caption{
  Redshift evolution of the ionizing background in our two models.  Solid curves correspond to the average inside cosmic HII regions, while dashed curves are the average over the whole volume (assuming a zero contribution from cosmic neutral regions).  Shaded regions denote the 1$\sigma$ scatter in the value of the UVB averaged on the scales of the HII regions.  Without requiring further tuning, the \smallHII\ model is consistent with current observations, also allowing room for an additional contribution to the UVB from AGN at lower redshifts (not accounted for in our models).
  }
\label{fig:gamma}
\vspace{-0.5\baselineskip}
\end{figure}

In Fig. \ref{fig:gamma} we show the corresponding redshift evolution of the ionizing background.   As cautioned in \citet{SM14},  our sub-grid models start to break down post overlap, as recombining systems become increasingly rare, hosted by large, poorly-understood overdensities.  Hence, we impose by hand a maximum mean free path of 75 Mpc, which is the cause of the flattening and lack of dispersion seen in the ionizing background for the \largeHII\ model post EoR at $z\lsim7$.  Even so, the ionizing background is too high in this model, especially given the fact that AGN could start dominating at $z\lsim6$ \citep{MH15}.  This tension with \lya\ forest estimates of the ionizing background at $z\lsim6$ could be ameliorated with a decreasing escape fraction of ionizing photons with decreasing redshift (e.g. \citealt{KF-G12, Khaire15, Smith16}).

On the other hand, the more extended, feedback-limited EoR predicted by \smallHII\ leads naturally to a slowly-evolving ionizing background and large opacity fluctuations, consistent with \lya\ forest observations (see also Fig. 10 in \citealt{SM14}; \citealt{AMT15}).  This model also leaves room for an AGN population dominating the ionizing background at $z\lsim6$ (e.g. \citealt{CF06, MH15}).  This picture is consistent with the theoretical considerations discussed in \S \ref{sec:sources}, {\it favoring the \smallHII\ model over the \largeHII\ model} (see also \citealt{CFG08, MCF15}).

\section{21-cm forecasts}
\label{sec:results}

\subsection{Generic trends}

\begin{figure*}
\vspace{-1\baselineskip}
{
\includegraphics[width=\textwidth]{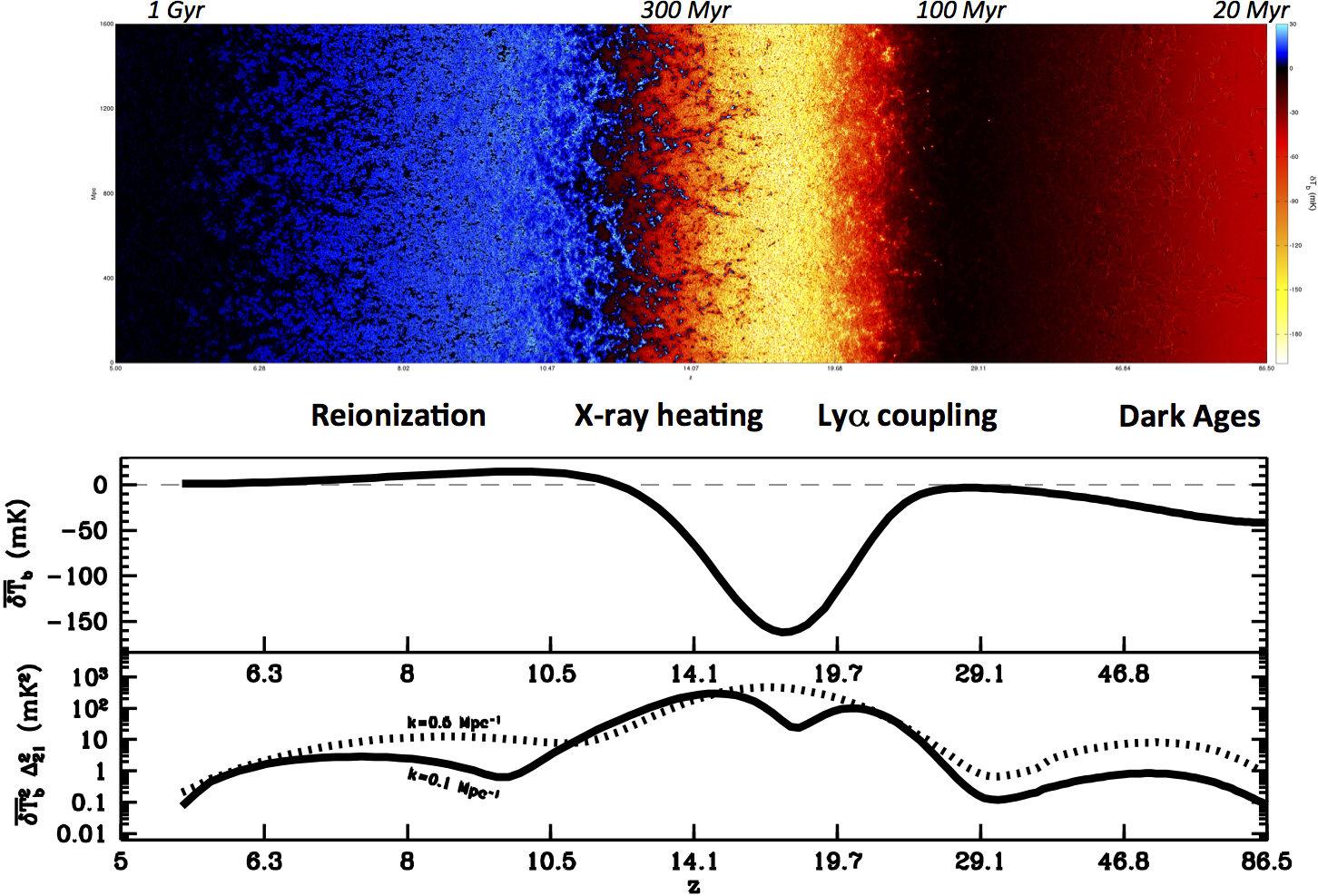}
}
\caption{
{\it Top:} light-cone strip corresponding to the \smallHII\ model. {\it Middle:} corresponding evolution of the global brightness temperature contrast. {\it Bottom:} corresponding evolution of the power spectrum amplitude at $k=0.1$ Mpc$^{-1}$ ({\it solid curve}) and $k=0.5$ Mpc$^{-1}$ ({\it dotted curve}). {\it Higher resolution version is available at http://homepage.sns.it/mesinger/EOS.html}
}
\label{fig:delT_strip_SmallHII}
\vspace{-0.5\baselineskip}
\end{figure*}

\begin{figure*}
\vspace{-1\baselineskip}
{
\includegraphics[width=\textwidth]{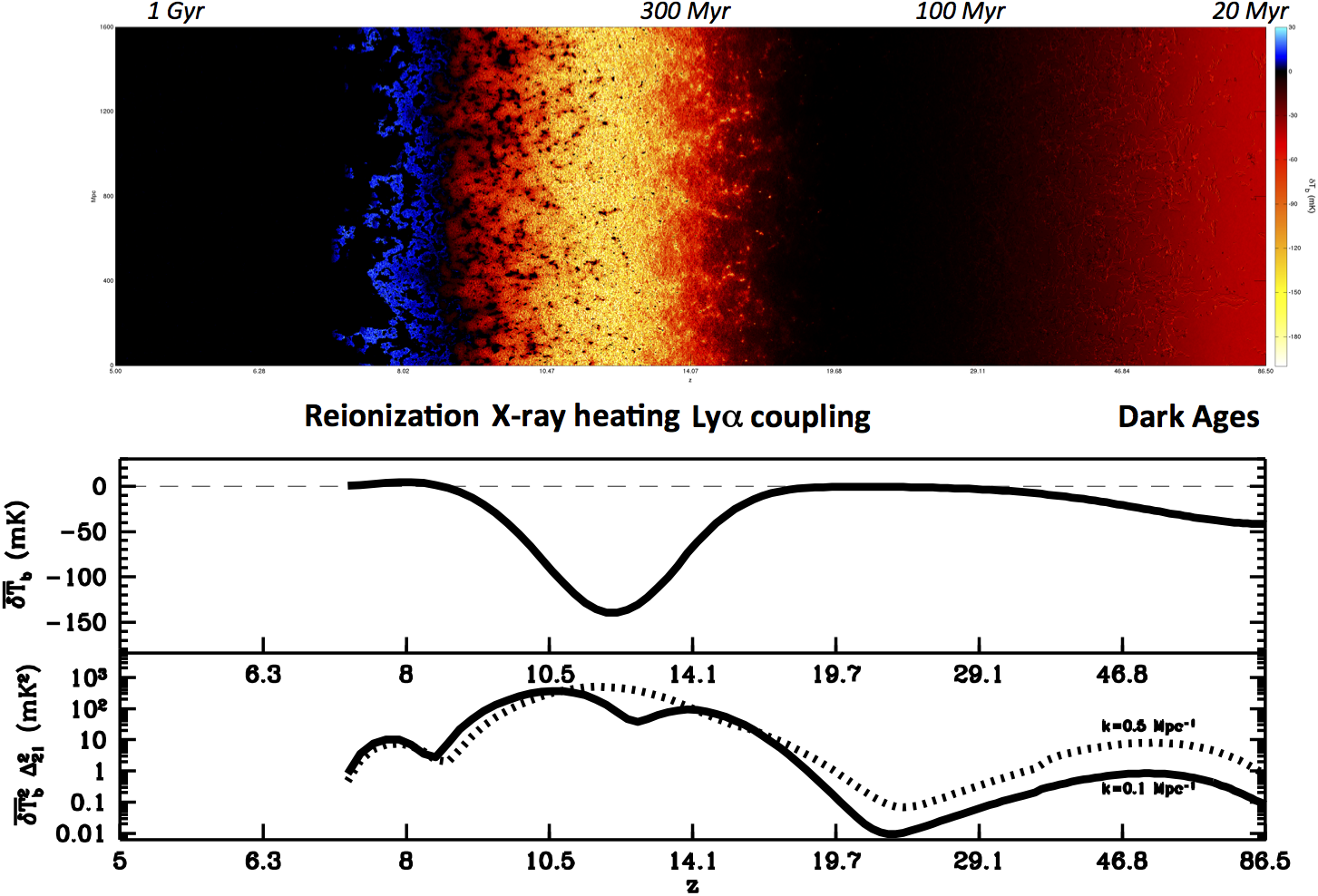}
}
\caption{
Same as Fig. \ref{fig:delT_strip_SmallHII}, but corresponding to the \largeHII\ model.  {\it Higher resolution version is available at http://homepage.sns.it/mesinger/EOS.html}
}
\label{fig:delT_strip_LargeHII}
\vspace{-0.5\baselineskip}
\end{figure*}

We now present the 21-cm signal resulting from our two models.  In Figures \ref{fig:delT_strip_SmallHII} and \ref{fig:delT_strip_LargeHII} we show the corresponding light cone strips ({\it top panels}), as well as the evolution of the global signal ({\it middle panels}) and the power spectrum amplitude (defined as $\bar{\delT}(z)^2  k^3 (2\pi^2 V)^{-1}  \langle|\delta_{\rm 21}({\bf k}, z)|^2\rangle_k$, with $\delta_{21}({\bf x}, z) \equiv \delT({\bf x}, z)/ \bar{\delT}(z) - 1$) at $k=$ 0.1 and 0.5 Mpc$^{-1}$ ({\it bottom panels}).  The cosmic epochs of \lya\ coupling, EoX, and the EoR are labeled in the figures.

The trends in these figures are consistent with the standard scenario for the evolution of the 21-cm signal, consisting of several epochs from high to low redshifts (e.g. \citealt{Furlanetto06, PF07, Baek10, Santos11, MFC11, PL12, MFS13, Furlanetto16}):
\begin{packed_enum}
\item {\bf Dark Ages - Collisional coupling}: The IGM is dense at high redshifts, so the spin temperature is uniformly collisionally coupled to the gas kinetic temperature, $T_K = T_S \lsim \Tcmb$. Following thermal decoupling from the CMB ($z\lsim$~200--300), the IGM cools adiabatically as $T_K \propto (1+z)^2$, faster than the CMB $\Tcmb \propto (1+z)$.  Thus $\avedelT$ is negative, and the power spectrum is determined by density fluctuations.  This part of the cosmic Dark Ages (corresponding to the red edge on the right side of Figures \ref{fig:delT_strip_SmallHII} and \ref{fig:delT_strip_LargeHII}; $50\lsim z \lsim200$) could serve as a clean probe of the matter power spectrum, if detected by some future experiment.
\item {\bf Dark Ages - Collisional decoupling}:  The IGM becomes less dense as the Universe expands.  The spin temperature starts to decouple from the kinetic temperature, and begins to approach the CMB temperature again, $T_K < T_S \lsim \Tcmb$.  Thus $\avedelT$ starts rising towards zero.  Decoupling from $\Tk$ occurs as a function of the local gas density, with underdense regions decoupling first. The matter power spectrum, modulated by the coupling strength, determines the 21-cm power spectrum. Eventually ($z\sim30$), all of the IGM is decoupled and there is little or no signal.  This epoch corresponds to the red$\rightarrow$black transition on the right edge of Figures \ref{fig:delT_strip_SmallHII} and \ref{fig:delT_strip_LargeHII}.    
\item {\bf Cosmic Dawn - WF (i.e. Ly$\alpha$) coupling}:  The first astrophysical sources turn on, and begin re-coupling $T_S$ and $T_K$, this time through the \lya\ background. $\avedelT$ becomes more negative, reaching values as low as $\avedelT\sim$ -100 -- -200 mK (depending on the offset of the WF coupling and X-ray heating epochs). This epoch, denoting the beginning of the Cosmic Dawn, corresponds to the black$\rightarrow$yellow transition in Figures \ref{fig:delT_strip_SmallHII} and \ref{fig:delT_strip_LargeHII}.\footnote{In our models the onset of \lya\ pumping is determined by the emergences of halos with the corresponding masses (see Fig. \ref{fig:Mhalo});
    Thus Cosmic Dawn starts earlier in the \smallHII\ model.  We caution however that even a small star formation density is sufficient to strongly couple $T_S$ and $T_K$ (e.g. \citealt{MO12}).  Given that the first stars are likely hosted by molecularly cooled halos, we expect the Cosmic Dawn to begin earlier than predicted in our models (which are motivated by theoretical and observational considerations at $z\sim$~6--10).}
\item{\bf Cosmic Dawn - X-ray heating}: The IGM is heated, with the spin temperature becoming increasingly coupled to the gas temperature, $T_K = T_S$ (see the bottom panel of Fig. \ref{fig:global}). As the gas temperature surpasses $\Tcmb$, the 21cm signal changes from absorption to emission, becoming insensitive to the actual value of $T_S$ (see eq. \ref{eq:delT}).  The power spectrum is expected to peak during this epoch, driven by the spatial fluctuations in the IGM temperature, which offer the largest dynamic range for the 21-cm signal (i.e. heated patches emit at $\delta T_b \sim 30$ mK, while cold patches are in absorption at $\delta T_b \sim$ -100 mK).
  This epoch corresponds to the yellow$\rightarrow$blue transition in Figures \ref{fig:delT_strip_SmallHII} and \ref{fig:delT_strip_LargeHII}.
\item {\bf Reionization}:  At around $z\sim8$, the Universe is undergoing reionization.  The power spectrum reaches a peak at roughly the midpoint of the EoR, driven by the order unity fluctuations in the neutral fraction.  The morphology of this process is sensitive to the nature and clustering of the dominant UV sources (e.g. \citealt{McQuinn07, GM15, Dixon15}).  The cosmic 21cm signal decreases, approaching zero.  This epoch corresponds to the blue$\rightarrow$black transition in the panels of  Figures \ref{fig:delT_strip_SmallHII} and \ref{fig:delT_strip_LargeHII}.
\end{packed_enum}

\begin{figure*}
\vspace{-1\baselineskip}
{
  \includegraphics[width=0.45\textwidth]{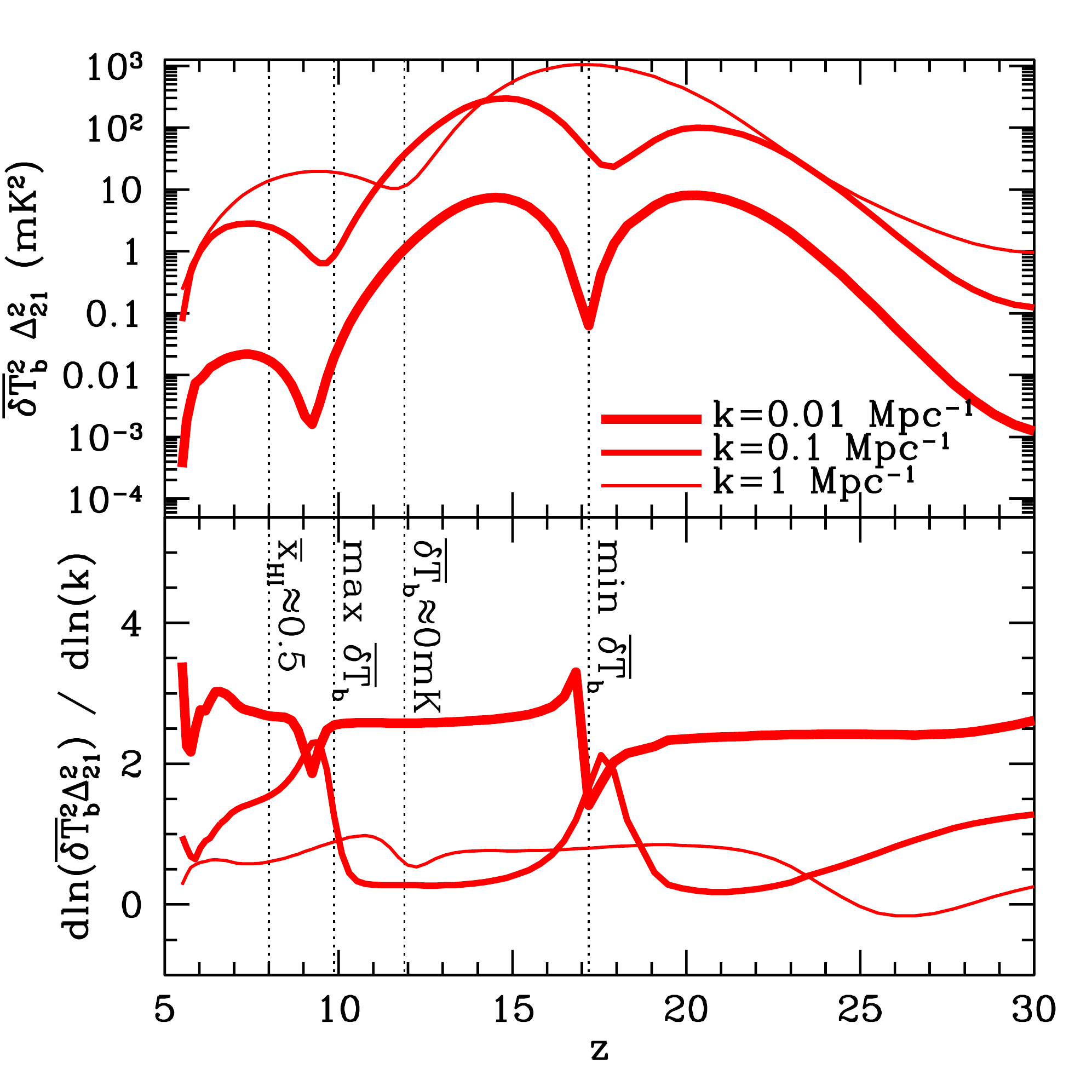}
  \includegraphics[width=0.45\textwidth]{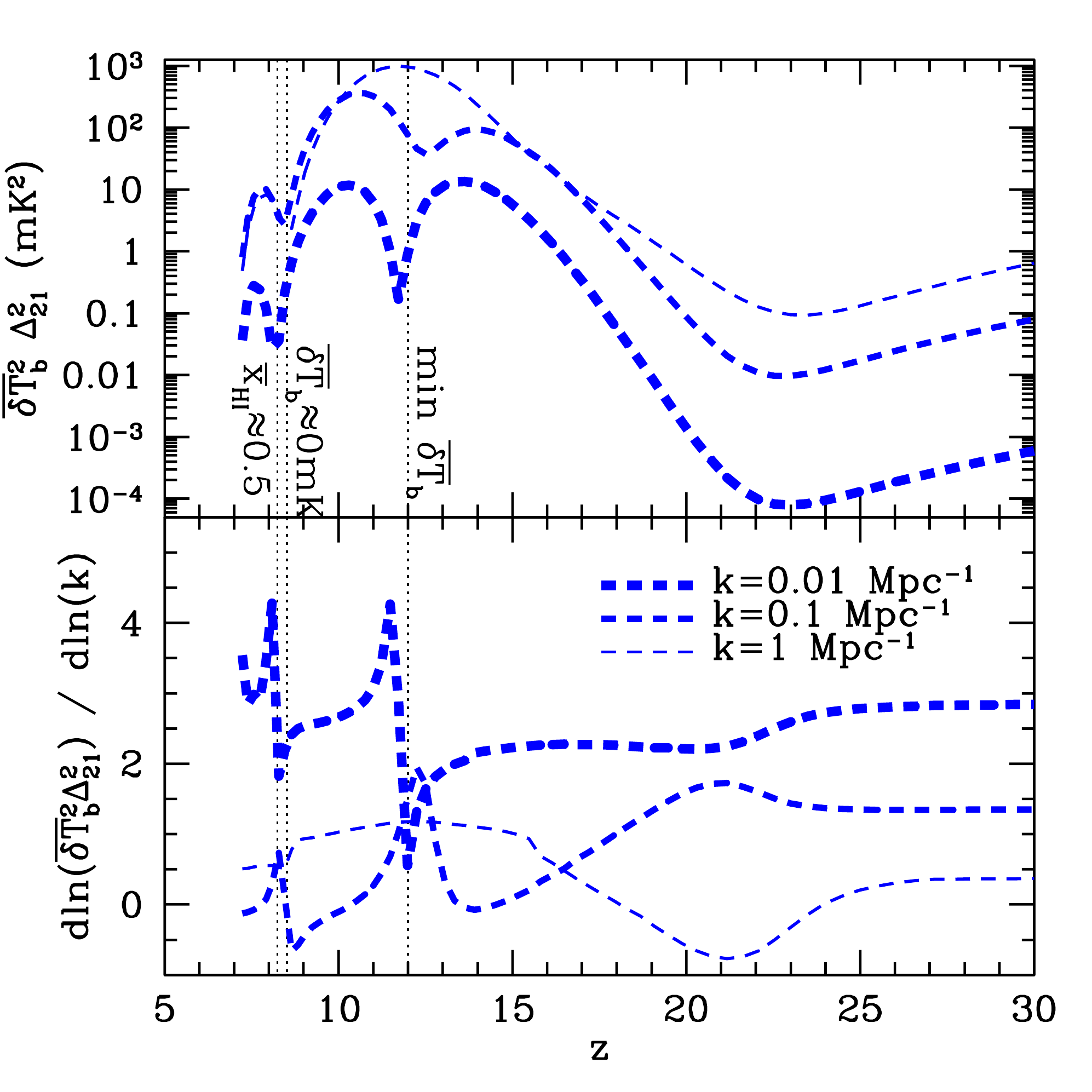}
}
\caption{
  Redshift evolution of the 21-cm power spectrum amplitude ({\it top panels}) and its derivative with respect to the logarithm of $k$ ({\it bottom panel}).  The \smallHII\ / \largeHII\ models are shown on the left / right.  The vertical dotted lines correspond to when $\avedelT$ is at a minimum, when it is zero, and when $\avenf=0$.  In the left panel where space allows, we also denote when $\avedelT$ is at a maximum.
  }
\label{fig:powervsz}
\vspace{-0.5\baselineskip}
\end{figure*}

\subsection{21-cm power spectrum}
\label{sec:PS}

We quantify these trends further in Fig. \ref{fig:powervsz}, which shows the evolution of the amplitude ({\it top panels}), derivative with $k$ ({\it bottom panels}).  The vertical dotted lines correspond to when $\avedelT$ is at its minimum, when it is zero, and when $\avenf=0.5$.  In the left panel, where space allows, we also denote when $\avedelT$ is at its maximum.  As noted in prior studies, the peaks and troughs of the evolution of the large-scale power demarcate milestones in the Universe's evolution.  The most notable difference between the two models is the relative timing of the epochs, which in the \largeHII\ model occur rapidly with significant overlap\footnote{As was already noted, in this model the global signal switches to emission very late, when $\avenf\sim0.5$. The resulting values of $\delta T_b \sim 0$ mK in cosmic HI regions strongly suppresses the contrast against the cosmic HII regions, and corresponding 21-cm power is much lower than expected with the common simplifying assumption of $\Ts \gg \Tcmb$}.

The {\it peaks of the large-scale power correspond roughly to the midpoints of the three astrophysical epochs: \lya\ pumping, EoX, EoR.}  Both models show this three peaked structure in the evolution of the large-scale power,
\footnote{We note that here we do not explore alternate scenarios for the heating epoch.  Exotic scenarios, such as heating by very hard, heavily obscured sources \citep{MFS13, FBV14} or dark matter annihilation \citep{EMF14}, result in a uniform heating which can dramatically suppress the peak in power associated with the heating epoch, as well as the trough between the heating and WF coupling epochs.}
driven by large-scale fluctuations in WF coupling, gas temperature, and the ionization fraction (from high to low redshift; e.g. \citealt{PF07, Baek10}).  Contrary to preliminary estimates assuming $\Ts \gg \Tcmb$ (e.g. \citealt{Lidz08, Friedrich11}), the EoR peak does not happen exactly at the mid-point of $\avenf=0.5$, but instead occurs afterwards.  This delay is especially notable in the \largeHII\ model, in which the heating and EoR epochs overlap strongly.  This is due to the $(1-\Tcmb/\Ts)^2$ contribution to the power spectrum from the mean brightness temperature, which contributes a factor of 0.9 (0.3) at the midpoint of the EoR in the \smallHII\ (\largeHII) models.  As more time passes after the EoR midpoint and X-ray sources continue to heat the cosmic neutral patches, the rise in $(1-\Tcmb/\Ts)^2$ more than compensates for the drop in $\Delta^2_{21}$, and so the peak in the power amplitude occurs at $\avenf < 0.5$.  We confirm that if $(1-\Tcmb/\Ts)^2$ is set to unity as is commonly done in the literature, we recover the result that the EoR power spectrum peaks at $\avenf\approx0.5$.

On the other hand the {\it troughs in the large-scale power evolution correspond roughly to the boundaries between these three epochs.}  As can be seen from the bottom panels of Fig. \ref{fig:powervsz}, they are marked by sudden changes in the slope of the power with $k$.  These are driven by the brief periods between the astrophysical epochs, when the cross-correlations in the brightness temperature components dominate the power (e.g. \citealt{Lidz08, PF07, MFS13}).  In the early stages of the EoR, the large-scale power drops as the densest patches close to galaxies are reionized, thus transitioning from being the strongest 21-cm emitters to having zero signal.  Likewise during the first stages of X-ray heating, these large-scale dense patches close to galaxies are the first to be heated, thus transitioning from being the strongest 21-cm absorbers (with the highest levels of WF coupling) to sourcing a much weaker emission signal.

\begin{figure}
\vspace{-1\baselineskip}
{
  \includegraphics[width=0.45\textwidth]{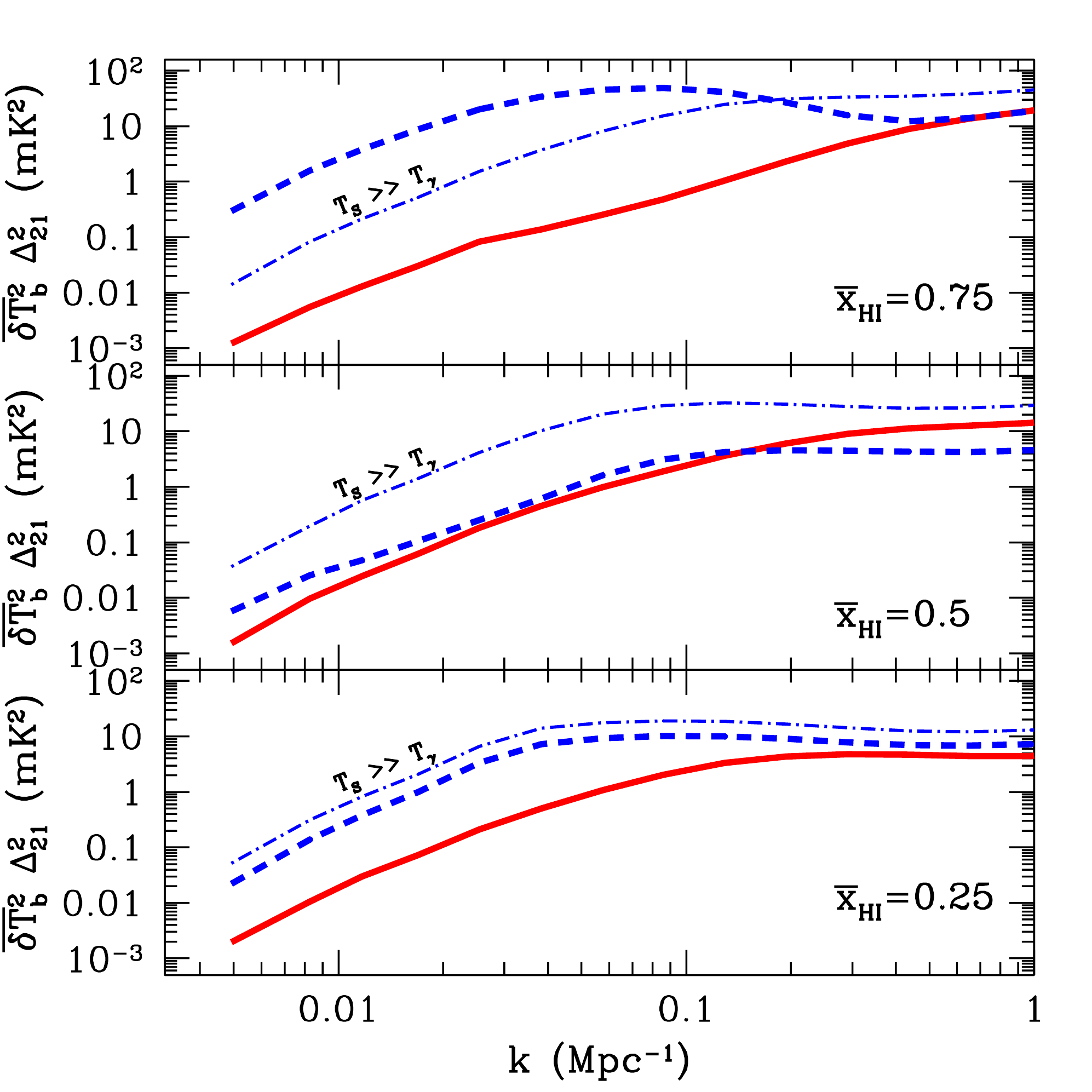}
}
\caption{
The 21-cm power spectrum during the EoR.  Solid red / Dashed blue curves correspond to the \smallHII\ / \largeHII\ models.  We also include a thin dot-dashed blue curve which corresponds to the \largeHII\ model, but with a saturated $\Ts \gg \Tcmb$, such as could be expected if the first galaxies formed in less massive halos than those hosting dominant reionizing sources or if the X-ray luminosity to star formation rate in the first galaxies is higher than observed in local galaxies.
}
\label{fig:powervsk}
\vspace{-0.5\baselineskip}
\end{figure}

Finally, in Fig. \ref{fig:powervsk} we show the power spectrum as a function of $k$, at various stages of the EoR: $\avenf \approx$ 0.75, 0.5, 0.25 ({\it top to bottom}).  In addition to our two fiducial models, we include as a reference a curve corresponding to the \largeHII\ model, but assuming saturated heating $\Ts \gg \Tcmb$.  As argued above, this is not unreasonable given that the galaxies responsible for X-ray heating could correspond to an earlier population hosted by smaller halos or have a significantly higher heating efficiency than we assume based on local galaxies (due to for example a higher HMXB fraction motivated by the lower metallically of the first galaxies; \citealt{Fragos12}).  Comparing the two  \largeHII\ curves at $\avenf \approx0.75$, we see that the 21-cm power in our fiducial model is still dominated by temperature fluctuations on large-scales.  In contrast, by $\avenf \approx 0.5$, the fiducial \largeHII\ model has a lower 21-cm power, due to the decreased contrast of HII and HI regions, the later of which are no longer very cold but are neither fully heated thus having $|\delT|\sim$ 0 mK (assuming the spin temperature has saturated would result in a factor of $\gsim$10 higher large scale power).

During the late stages of the EoR, the spin temperature saturates, allowing the 21-cm power spectrum to be dominated primarily by the reionization morphology.  In this regime, the \largeHII\ model has a factor of $\gsim10$ more large-scale power than the \smallHII\ model.  This is due to two effects: (i) the higher bias of the typical reionizing sources in the \largeHII\ model (resulting in a factor of $\sim$2--4 higher power; e.g. \citealt{McQuinn07, GM15}); (ii) sub-grid recombinations and photo-heating feedback which combine to suppress large-scale power by factors of $\sim$2--3 in models with an extended reionization driven by galaxies close to the atomic cooling threshold \citep{SM14}.

\subsection{Detectability}
\label{sec:detect}

\begin{table*}
\begin{tabular}{@{}lccccccc}
\hline
Parameter & LOFAR & PAPER & MWA 128T & HERA & SKA & SKA (with substations)\\
\hline
Telescope antennae & 48 & 132 &128 & 331 & 564 & 3384 \\
Diameter (m) & 30.8 & 3.4 & 5.2 & 14 & 30 & 10\\
Collecting Area (m$^2$) & 35\,762 & 1\,198 & 2\,752 & 50\,953 & 398\,668 & 265\,779\\
$T_{\rm rec}$ (K) & 140 & 100 & 50 & 100 & 0.1T$_{\rm sky}$ + 40 & 0.1T$_{\rm sky}$ + 40 \\
Bandwidth (MHz) & 8 & 8 & 8 & 8 & 8 & 8\\
Integration time (hrs) & 1000 & 1000 & 1000 & 1000 & 10-1000 & 10-1000\\
Observing frequency (MHz) & 110-250 & 100-200 & 75-250 & 50-250 & 50-250 & 50-250 \\
\hline
\end{tabular}
\caption{Summary of telescope parameters we use to compute sensitivity profiles (see text for further details).}
\label{tab:TelescopeParams}
\end{table*}

We now present signal-to-noise (S/N) forecasts for current and upcoming interferometers.  We compute the telescope noise profiles using a parallelized version of the publicly available code \sense\footnote{https://github.com/jpober/21cmSense}\citep{Pober13, Pober14}.  We refer the reader to these works for more details, and here we briefly summarize the main assumptions.

The thermal noise PS is computed at each $uv$-cell  according to the following \citep[e.g.][]{Morales05,McQuinn06},
\begin{eqnarray} \label{eq:NoisePS}
\Delta^{2}_{\rm N}(k) \approx X^{2}Y\frac{k^{3}}{2\upi^{2}}\frac{\Omega^{\prime}}{2t}T^{2}_{\rm sys},
\end{eqnarray} 
where $X^{2}Y$ is a cosmological conversion factor between observing bandwidth (we take a constant bandwidth of 8MHz throughout), frequency and comoving volume, $\Omega^{\prime}$ is a beam-dependent factor derived in \citet{Parsons14}, $t$ is the total time spent by all baselines within a particular $k$ mode for a single pointing, and $T_{\rm sys}$ is the system temperature, the sum of the receiver temperature, $T_{\rm rec}$, and the sky temperature $T_{\rm sky}= 60\left(\frac{\nu}{300~{\rm MHz}}\right)^{-2.55}~{\rm K}$ \citep{TMS07}. 

The total noise PS is then computed by performing an inverse-weighted summation over all individual modes, $i$, \citep[e.g.][]{McQuinn06}, combining the contribution from the sample variance and the thermal noise  from equation~\ref{eq:NoisePS}):
\begin{eqnarray} \label{eq:T+S}
\delta\Delta^{2}_{\rm T+S}(k) = \left(\sum_{i}\frac{1}{[\Delta^{2}_{{\rm N},i}(k) + \Delta^{2}_{21}(k)]^{2}}\right)^{-1/2}.
\end{eqnarray}
Here $\delta\Delta^{2}_{\rm T+S}(k)$ is the total uncertainty from thermal noise and sample variance in a given $k$-mode and $\Delta^{2}_{21}(k)$ is the expectation value for the cosmological power spectrum from our models.  We assume that the sample variance of $\Delta^{2}_{21}(k)$ has a Poisson distribution, which is a good approximation on large-scales \citep{Mondal15}.

We take a fiducial integration time of 1000h.  For tracked-scan instruments (MWA, LOFAR and SKA1-low) we assume a 1000h integration on a single pointing.  We also present estimates for the planned shallow, wide, 100$\times$10h SKA1-low survey, which for a fixed total observing time of 1000h increases the sky area covered (reduced sample variance) at the expense of increasing thermal noise. For all observations we assume that all baselines can be correlated.  We take the antennae positions and specifications from the following: LOFAR \citep{vanHaarlem13}, PAPER \citep{Parsons10,Parsons12}, MWA \citep{Tingay13}, HERA \citep{Beardsley15}, SKA1-low [the Flower (F.) design from http://astronomers.skatelescope.org/wp-content/uploads/2015/11/SKA1-Low-Configuration\_V4a.pdf].  For SKA1-Low, we include an additional telescope configuration, Flower with Substations (F.S.), consisting of 10m substations, optimistically assuming all substations can be correlated; we note however that the current default design does not include substations.
The parameter choices used in our telescope noise models are summarized in Table \ref{tab:TelescopeParams}.

\begin{figure*}
\vspace{-1\baselineskip}
  \includegraphics[width=0.4\textwidth]{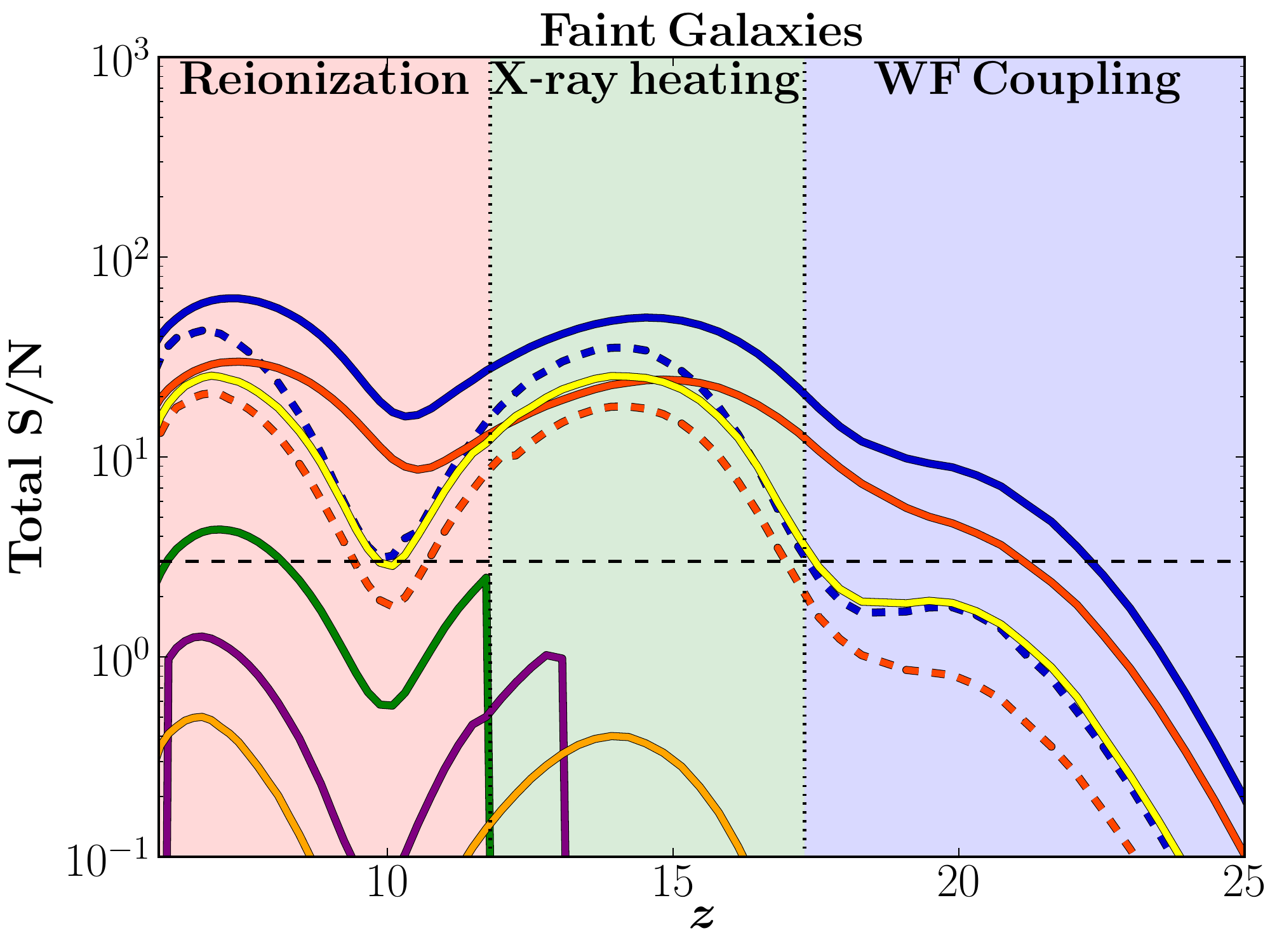}
  \includegraphics[width=0.4\textwidth]{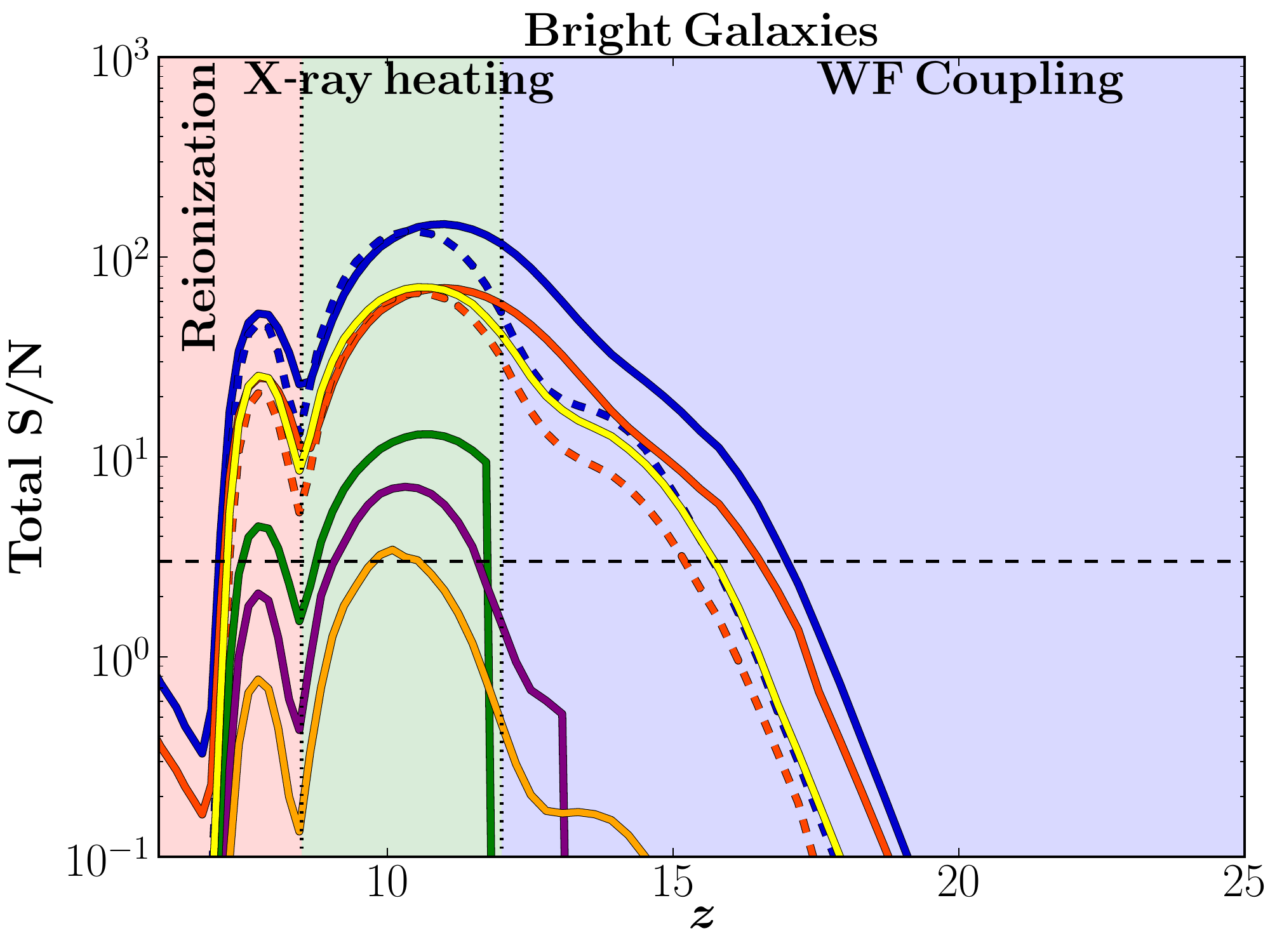}
  \hspace{0.3cm}
  \includegraphics[trim = 0.5cm -4.5cm 0cm 0cm, width=0.15\textwidth]{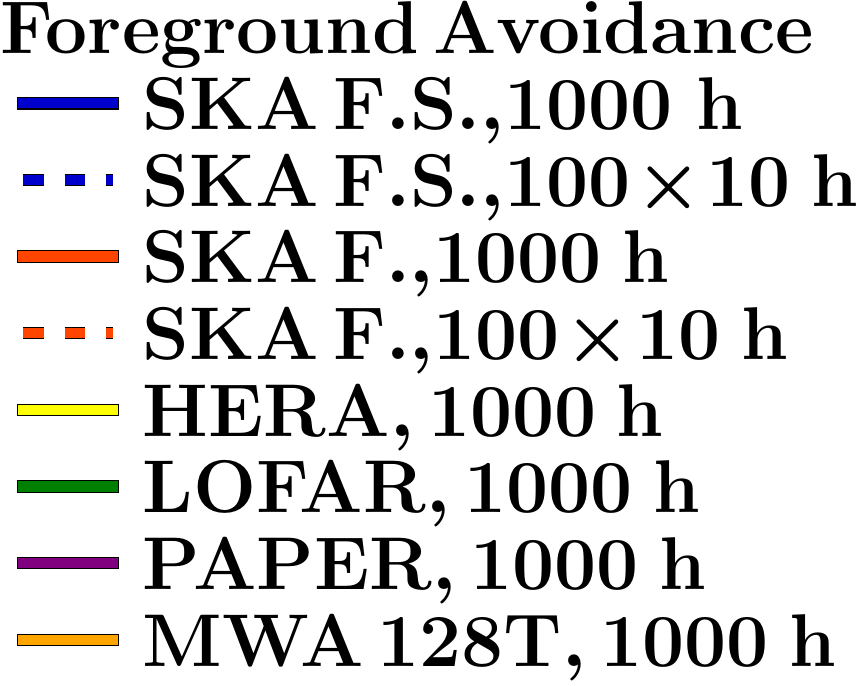}
\caption{
  Redshift evolution of the total (summed over all $k$ bins) S/N, for various interferometers and survey strategies.  The left (right) panel corresponds to \smallHII\ (\largeHII).  Here we assume a ``moderate'' scenario for foreground contamination, excising the foreground-contaminated ``wedge'' and assuming no contamination beyond the wedge.  The thin horizontal line denotes S/N = 3.
  The astrophysical epochs of WF coupling, X-ray heating, and EoR are crudely demarcated based on the global signal: vertical boundaries correspond to $\avedelT(z)=0$ mK (EoR / Heating) and $\avedelT(z) = $ min[$\avedelT$] (Heating / WF coupling; see \S \ref{sec:PS} for a more nuanced discussion of the astrophysical epochs and their overlap).
}
\label{fig:SNvsz_moderate}
\vspace{-0.5\baselineskip}
\end{figure*}

In Fig. \ref{fig:SNvsz_moderate} we present estimates of the total, integrated S/N for our simulations, assuming a ``moderate'' scenario for foreground contamination (see \citealt{Pober14}).  Specifically, when computing the noise, we excise the contribution of modes confined within a 2D $k$-space ``wedge'', which is defined to extend $\Delta k_{\parallel} = 0.1 h$~Mpc$^{-1}$ beyond the horizon limit. We then obtain the total S/N by summing over all modes.

The left panel of the figure corresponds to our \smallHII\ model.  We recover the result of \citet{ME-WH14}: that in fiducial scenarios, the EoR and EoX can be detected with comparable S/N.  This is due to the fact that the higher signal during X-ray heating (e.g. Fig. \ref{fig:delT_strip_SmallHII}), compensates for the increase in noise going to lower frequencies (we note however that additional systematics might become more difficult at lower frequencies).  This is no longer the case in the \largeHII\ scenario, since the EoX occurs at lower redshifts, where the noise is lower, allowing it to be detected with the highest S/N.
Indeed, current PAPER-64 upper limits at $z\sim11$ of $\avedelT^2\Delta_{21}^2(k\sim0.1$Mpc$^{-1}) \lsim 500$ mK$^2$ (Kolopanis et al., in prep) are only a factor of $\sim$2 away from the X-ray heating signal in the \largeHII\ model.

We note that HERA and the default SKA1-low flower design (F.) will be able to detect the peaks of the EoR and EoX at roughly the same, cosmic variance limited S/N.  However, a single deep 1000h pointing with SKA1-low performs better than either HERA or the wide-field 100x10h SKA1-low survey when the signal is low and the detection is thermal noise limited.  This is consistent with the results of \citet{GM15} and \citet{GMK15} who find that the wide area surveys, available with HERA's drift scan and the SKA1-low 100x10h fields, can deliver comparable or even better constraints on EoR astrophysics from the power spectrum.  These constraints result from the ability of the cosmic-variance limited low-$k$ modes to best discriminate between EoR models.
However, deep integrations which improve on the thermal noise component (like the planned single 1000h pointing with SKA1-low) can deliver higher overall S/N, allowing for (i) higher quality imaging of the EoR; (ii) statistical detections at the highest redshifts.

The overall winner is the 1000h single field integration with the SKA1-low flower design with substations (F.S.).  This design however is optimistic, as it assumes that all of the 3384 baselines can be correlated (though in practice the S/N is determined only by the inner $\sim$1400 baselines, likely a relatively achievable number for GPU correlators in the near future).  The current default design, SKA1-low flower (F.), is a factor of two worse in S/N.

\begin{figure*}
\vspace{-1\baselineskip}
  \includegraphics[width=0.4\textwidth]{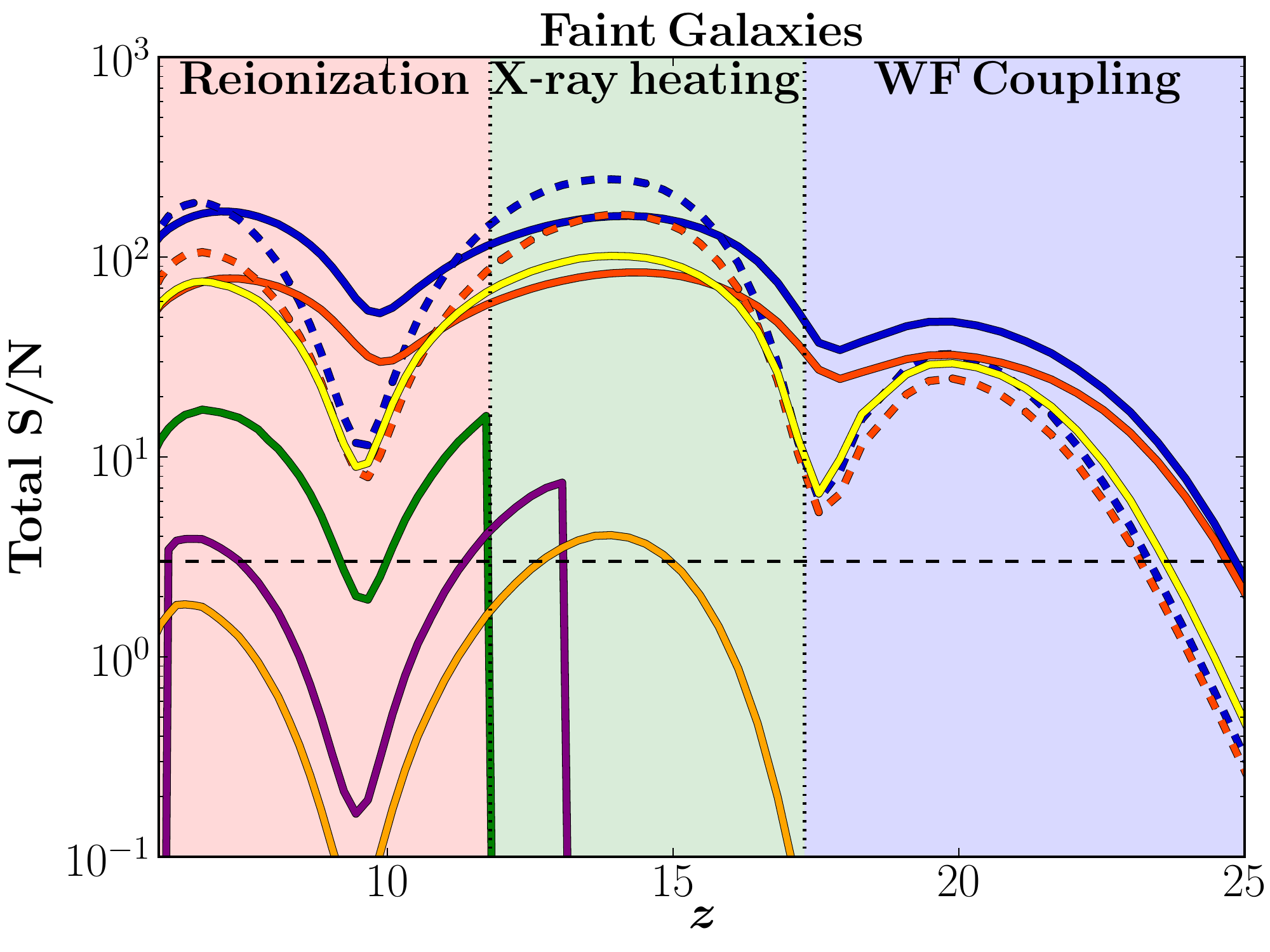}
  \includegraphics[width=0.4\textwidth]{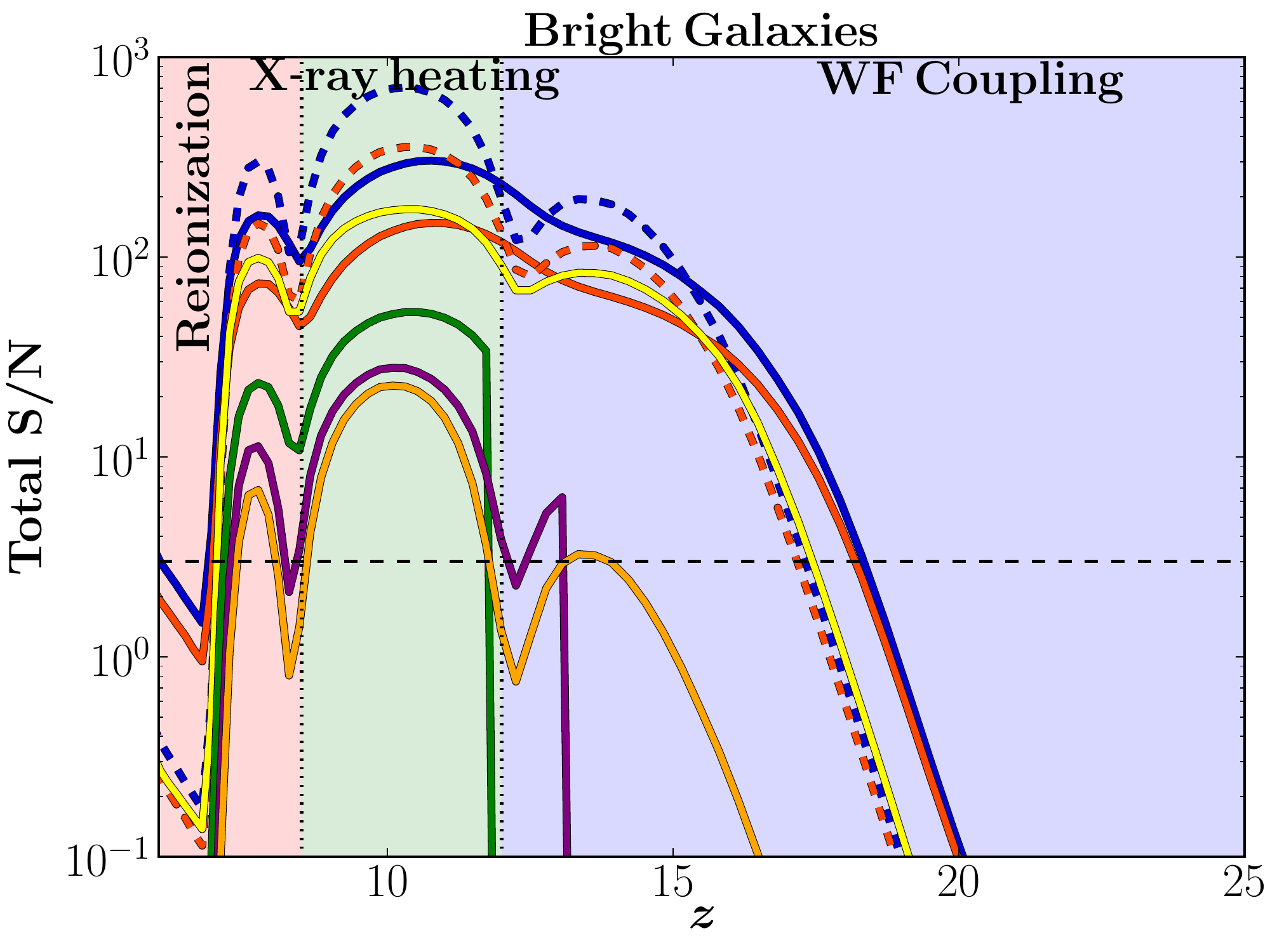}
  \hspace{0.3cm}
  \includegraphics[trim = 0.5cm -4.5cm 0cm 0cm, width=0.15\textwidth]{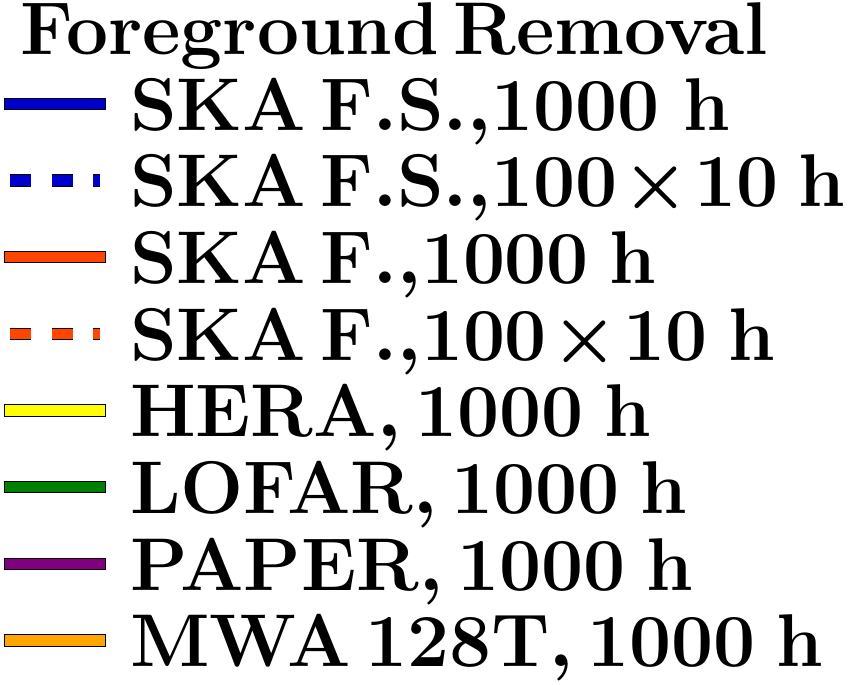}
\caption{
Same as Fig. \ref{fig:SNvsz_moderate}, but assuming an ``optimistic'' scenario in which all foregrounds can be successfully cleaned.
}
\label{fig:SNvsz_optimistic}
\vspace{-0.5\baselineskip}
\end{figure*}

Finally in Fig. \ref{fig:SNvsz_optimistic}, we show the analogous forecasts, but now assuming an ``optimistic'' foreground scenario, in which all foregrounds can be removed, even those in the ``wedge''.  In this maximally optimistic scenario, all telescopes should be able to achieve a statistical detection after 1000h.  Moreover, the above-mentioned trends of deep vs wide for SKA1-low observations are reversed: with perfect foreground cleaning, the cosmic-variance limits at the foreground-contaminated low-$k$ modes can be reached in the 100x10h survey, allowing for a higher overall S/N than the single 1000h pointing.





\section{Conclusions}
\label{sec:conc}

We introduce the 2016 data release of the Evolution of 21-cm Structure (EOS) project.  The goal of EOS is to provide periodic public releases of large, physics-rich simulations of the cosmic 21-cm signal.  Although not intended as a replacement for astrophysical parameter studies, EOS releases can be used to aid ongoing efforts in the construction of realistic data-analysis pipelines for 21-cm interferometers. 

In this release, we provide 1.6 Gpc ($\sim$9 deg) simulations, on a 1024$^3$ grid, over the redshift range $5\lsim z \lsim 200$.  We include two extreme models which approximately bracket the expected contribution from undetected galaxies: \smallHII\ and \largeHII.  In \smallHII, the EoR is driven primarily by galaxies close to the atomic cooling threshold, which are vulnerable to photo-heating suppression of their gas content.  As a result, the dominant ionizing population resides in halos of mass $\sim10^{8.5-9} \Msun$, with this value showing a large spatial scatter and rapid evolution at $z\lsim7$.  In \largeHII, the EoR is driven primarily by galaxies inside $\sim10^{10} \Msun$ halos, roughly corresponding to the faintest, currently observed LBGs at high redshifts.
Both simulations have their ionizing efficiency (e.g. escape fraction) calibrated to result in a Thompson scattering optical depth consistent with the latest measurement from the Planck satellite \citep{Planck15}.  We stress that the \smallHII\ model is likely closer to the true Universe, based on both theoretical and observational arguments.

Consistent with previous studies, in \smallHII, the EoR evolves slowly, and is characterized by small ionized structures.  On the other hand, in \largeHII, the EoR evolves rapidly, with large, more isolated HII regions.  These extremes suggest that the duration of reionization (defined as a change in the mean neutral fraction from 0.9 to 0.1) should be between  $2.7 \lsim \Delta z_{\rm re} \lsim 5.7$.  Moreover, we find that the large-scale 21-cm power during the advanced EoR stages can be different by up to a factor of $\sim10$.  This difference has a comparable contribution from: (i) the typical bias of sources in the two extreme models; (ii) a more efficient negative feedback in models with an extended reionization driven by faint galaxies.

We confirm that the peaks in the redshift evolution of the large-scale power correspond roughly to the midpoints of the three astrophysical epochs: (i) \lya\ pumping, (ii) X-ray heating, (iii) EoR.  However, the EoR peak occurs somewhat after the midpoint, due to the overlap between (ii) and (iii).  On the other hand, the throughs in the large-scale power evolution roughly correspond to the boundaries between (i)--(iii).

We find that HERA and SKA1-low should detect the EoR and epoch of X-ray heating with a 1000h observation at a S/N of few--hundreds, depending primarily on the success of foreground cleaning.  In the maximally optimistic assumption that all foregrounds can be removed, all first generation instruments could have a statistical detection (S/N $\gsim$ 3) after 1000h.

Our 21-cm power spectra, data cubes and visualizations are available at http://homepage.sns.it/mesinger/EOS.html.  This page will be updated with future releases, as well as scripts to convert the format of the output, facilitating its inclusion in data pipelines.


\section*{Acknowledgements}

This project has received funding from the European Research Council (ERC) under the European Union's Horizon 2020 research and innovation programme (grant agreement No 638809 -- AIDA).

\bibliographystyle{mn2e}
\bibliography{ms}

\begin{thebibliography}{}

\bibitem[\protect\citeauthoryear{{Atek} et~al.,}{{Atek}  et~al.}{2015}]{Atek15}
{Atek} H.,  et~al., 2015, \apj, 814, 69

\bibitem[\protect\citeauthoryear{{Baek}, {Semelin}, {Di Matteo}, {Revaz} \&
  {Combes}}{{Baek} et~al.}{2010}]{Baek10}
{Baek} S.,  {Semelin} B.,  {Di Matteo} P.,  {Revaz} Y.,    {Combes} F.,  2010,
  \aap, 523, A4

\bibitem[\protect\citeauthoryear{{Barkana} \& {Loeb}}{{Barkana} \&
  {Loeb}}{2001}]{BL01}
{Barkana} R.,  {Loeb} A.,  2001, \physrep, 349, 125

\bibitem[\protect\citeauthoryear{{Barkana} \& {Loeb}}{{Barkana} \&
  {Loeb}}{2004}]{BL04}
{Barkana} R.,  {Loeb} A.,  2004, \apj, 609, 474

\bibitem[\protect\citeauthoryear{{Barkana} \& {Loeb}}{{Barkana} \&
  {Loeb}}{2005}]{BL05_WF}
{Barkana} R.,  {Loeb} A.,  2005, \apj, 626, 1

\bibitem[\protect\citeauthoryear{{Barone-Nugent} et~al.,}{{Barone-Nugent}
  et~al.}{2014}]{Barone-Nugent14}
{Barone-Nugent} R.~L.,  et~al., 2014, \apj, 793, 17

\bibitem[\protect\citeauthoryear{{Beardsley}, {Morales}, {Lidz}, {Malloy} \&
  {Sutter}}{{Beardsley} et~al.}{2015}]{Beardsley15}
{Beardsley} A.~P.,  {Morales} M.~F.,  {Lidz} A.,  {Malloy} M.,    {Sutter}
  P.~M.,  2015, \apj, 800, 128

\bibitem[\protect\citeauthoryear{{Choudhury} \& {Ferrara}}{{Choudhury} \&
  {Ferrara}}{2006}]{CF06}
{Choudhury} T.~R.,  {Ferrara} A.,  2006, \mnras, 371, L55

\bibitem[\protect\citeauthoryear{{Choudhury}, {Ferrara} \&
  {Gallerani}}{{Choudhury} et~al.}{2008}]{CFG08}
{Choudhury} T.~R.,  {Ferrara} A.,    {Gallerani} S.,  2008, \mnras, 385, L58

\bibitem[\protect\citeauthoryear{{Cooke}, {Ryan-Weber}, {Garel} \&
  {D{\'{\i}}az}}{{Cooke} et~al.}{2014}]{Cooke14}
{Cooke} J.,  {Ryan-Weber} E.~V.,  {Garel} T.,    {D{\'{\i}}az} C.~G.,  2014,
  \mnras, 441, 837

\bibitem[\protect\citeauthoryear{{D'Aloisio}, {McQuinn} \& {Trac}}{{D'Aloisio}
  et~al.}{2015}]{AMT15}
{D'Aloisio} A.,  {McQuinn} M.~J.,    {Trac} H.,  2015, ArXiv
  e-prints:1509.02523

\bibitem[\protect\citeauthoryear{{Dijkstra}, {Gilfanov}, {Loeb} \&
  {Sunyaev}}{{Dijkstra} et~al.}{2012}]{Dijkstra12}
{Dijkstra} M.,  {Gilfanov} M.,  {Loeb} A.,    {Sunyaev} R.,  2012, \mnras, 421,
  213

\bibitem[\protect\citeauthoryear{{Dixon}, {Iliev}, {Mellema}, {Ahn} \&
  {Shapiro}}{{Dixon} et~al.}{2015}]{Dixon15}
{Dixon} K.~L.,  {Iliev} I.~T.,  {Mellema} G.,  {Ahn} K.,    {Shapiro} P.~R.,
  2015, ArXiv e-prints:1512.03836

\bibitem[\protect\citeauthoryear{{Evoli}, {Mesinger} \& {Ferrara}}{{Evoli}
  et~al.}{2014}]{EMF14}
{Evoli} C.,  {Mesinger} A.,    {Ferrara} A.,  2014, \jcap, 11, 024

\bibitem[\protect\citeauthoryear{{Ewall-Wice}, {Hewitt}, {Mesinger}, {Dillon},
  {Liu} \& {Pober}}{{Ewall-Wice} et~al.}{2015}]{EW15}
{Ewall-Wice} A.,  {Hewitt} J.,  {Mesinger} A.,  {Dillon} J.~S.,  {Liu} A.,
  {Pober} J.,  2015, ArXiv e-prints:1511.04101

\bibitem[\protect\citeauthoryear{{Fialkov}, {Barkana} \& {Visbal}}{{Fialkov}
  et~al.}{2014}]{FBV14}
{Fialkov} A.,  {Barkana} R.,    {Visbal} E.,  2014, \nat, 506, 197

\bibitem[\protect\citeauthoryear{{Fialkov}, {Barkana}, {Visbal},
  {Tseliakhovich} \& {Hirata}}{{Fialkov} et~al.}{2013}]{Fialkov13}
{Fialkov} A.,  {Barkana} R.,  {Visbal} E.,  {Tseliakhovich} D.,    {Hirata}
  C.~M.,  2013, \mnras, 432, 2909

\bibitem[\protect\citeauthoryear{{Field}}{{Field}}{1958}]{Field58}
{Field} G.~B.,  1958, Proceedings of the Institute of Radio Engineers, 46, 240

\bibitem[\protect\citeauthoryear{Fragos et~al.,}{Fragos
  et~al.}{2013}]{Fragos12}
Fragos T.,  et~al., 2013, \apj, 764, 41

\bibitem[\protect\citeauthoryear{{Friedrich}, {Mellema}, {Alvarez}, {Shapiro}
  \& {Iliev}}{{Friedrich} et~al.}{2011}]{Friedrich11}
{Friedrich} M.~M.,  {Mellema} G.,  {Alvarez} M.~A.,  {Shapiro} P.~R.,
  {Iliev} I.~T.,  2011, \mnras, 413, 1353

\bibitem[\protect\citeauthoryear{{Furlanetto}}{{Furlanetto}}{2006}]{Furlanetto06}
{Furlanetto} S.~R.,  2006, \mnras, 371, 867

\bibitem[\protect\citeauthoryear{{Furlanetto}}{{Furlanetto}}{2016}]{Furlanetto16}
{Furlanetto} S.~R.,  2016, in {Mesinger} A.,  ed., Astrophysics and Space
  Science Library Vol.~423 of Astrophysics and Space Science Library, {The
  21-cm Line as a Probe of Reionization}.
p.~247

\bibitem[\protect\citeauthoryear{{Furlanetto}, {Oh} \& {Briggs}}{{Furlanetto}
  et~al.}{2006}]{FOB06}
{Furlanetto} S.~R.,  {Oh} S.~P.,    {Briggs} F.~H.,  2006, \physrep, 433, 181

\bibitem[\protect\citeauthoryear{{Furlanetto} \& {Stoever}}{{Furlanetto} \&
  {Stoever}}{2010}]{FS10}
{Furlanetto} S.~R.,  {Stoever} S.~J.,  2010, \mnras, 404, 1869

\bibitem[\protect\citeauthoryear{{Furlanetto}, {Zaldarriaga} \&
  {Hernquist}}{{Furlanetto} et~al.}{2004}]{FZH04}
{Furlanetto} S.~R.,  {Zaldarriaga} M.,    {Hernquist} L.,  2004, \apj, 613, 1

\bibitem[\protect\citeauthoryear{{Geil}, {Mutch}, {Poole}, {Angel}, {Duffy},
  {Mesinger} \& {Wyithe}}{{Geil} et~al.}{2015}]{Geil15}
{Geil} P.~M.,  {Mutch} S.~J.,  {Poole} G.~B.,  {Angel} P.~W.,  {Duffy} A.~R.,
  {Mesinger} A.,    {Wyithe} J.~S.~B.,  2015, ArXiv e-prints

\bibitem[\protect\citeauthoryear{{Greig} \& {Mesinger}}{{Greig} \&
  {Mesinger}}{2015}]{GM15}
{Greig} B.,  {Mesinger} A.,  2015, \mnras, 449, 4246

\bibitem[\protect\citeauthoryear{{Greig}, {Mesinger} \& {Koopmans}}{{Greig}
  et~al.}{2015}]{GMK15}
{Greig} B.,  {Mesinger} A.,    {Koopmans} L.~V.~E.,  2015, ArXiv
  e-prints:1509.03312

\bibitem[\protect\citeauthoryear{{Haiman}, {Rees} \& {Loeb}}{{Haiman}
  et~al.}{1997}]{HRL97}
{Haiman} Z.,  {Rees} M.~J.,    {Loeb} A.,  1997, \apj, 484, 985

\bibitem[\protect\citeauthoryear{{Hirata}}{{Hirata}}{2006}]{Hirata06}
{Hirata} C.~M.,  2006, \mnras, 367, 259

\bibitem[\protect\citeauthoryear{{Holzbauer} \& {Furlanetto}}{{Holzbauer} \&
  {Furlanetto}}{2012}]{HF12}
{Holzbauer} L.~N.,  {Furlanetto} S.~R.,  2012, \mnras, 419, 718

\bibitem[\protect\citeauthoryear{{Hui} \& {Gnedin}}{{Hui} \&
  {Gnedin}}{1997}]{HG97}
{Hui} L.,  {Gnedin} N.~Y.,  1997, \mnras, 292, 27

\bibitem[\protect\citeauthoryear{{Khaire}, {Srianand}, {Choudhury} \&
  {Gaikwad}}{{Khaire} et~al.}{2015}]{Khaire15}
{Khaire} V.,  {Srianand} R.,  {Choudhury} T.~R.,    {Gaikwad} P.,  2015, ArXiv
  e-prints:1510.04700

\bibitem[\protect\citeauthoryear{{Koopmans} et~al.,}{{Koopmans}
  et~al.}{2015}]{Koopmans15}
{Koopmans} L.,  et~al., 2015, Advancing Astrophysics with the Square Kilometre
  Array (AASKA14), p.~1

\bibitem[\protect\citeauthoryear{{Kuhlen} \& {Faucher-Giguere}}{{Kuhlen} \&
  {Faucher-Giguere}}{2012}]{KF-G12}
{Kuhlen} M.,  {Faucher-Giguere} C.-A.,  2012, \mnras, 423, 862

\bibitem[\protect\citeauthoryear{{Lidz}, {Zahn}, {McQuinn}, {Zaldarriaga} \&
  {Hernquist}}{{Lidz} et~al.}{2008}]{Lidz08}
{Lidz} A.,  {Zahn} O.,  {McQuinn} M.,  {Zaldarriaga} M.,    {Hernquist} L.,
  2008, \apj, 680, 962

\bibitem[\protect\citeauthoryear{{Madau} \& {Haardt}}{{Madau} \&
  {Haardt}}{2015}]{MH15}
{Madau} P.,  {Haardt} F.,  2015, \apjl, 813, L8

\bibitem[\protect\citeauthoryear{{McQuinn}}{{McQuinn}}{2012}]{McQuinn12}
{McQuinn} M.,  2012, \mnras, 426, 1349

\bibitem[\protect\citeauthoryear{{McQuinn}, {Lidz}, {Zahn}, {Dutta},
  {Hernquist} \& {Zaldarriaga}}{{McQuinn} et~al.}{2007}]{McQuinn07}
{McQuinn} M.,  {Lidz} A.,  {Zahn} O.,  {Dutta} S.,  {Hernquist} L.,
  {Zaldarriaga} M.,  2007, \mnras, 377, 1043

\bibitem[\protect\citeauthoryear{{McQuinn} \& {O'Leary}}{{McQuinn} \&
  {O'Leary}}{2012}]{MO12}
{McQuinn} M.,  {O'Leary} R.~M.,  2012, \apj, 760, 3

\bibitem[\protect\citeauthoryear{{McQuinn}, {Zahn}, {Zaldarriaga}, {Hernquist}
  \& {Furlanetto}}{{McQuinn} et~al.}{2006}]{McQuinn06}
{McQuinn} M.,  {Zahn} O.,  {Zaldarriaga} M.,  {Hernquist} L.,    {Furlanetto}
  S.~R.,  2006, \apj, 653, 815

\bibitem[\protect\citeauthoryear{{Mesinger} \& {Dijkstra}}{{Mesinger} \&
  {Dijkstra}}{2008}]{MD08}
{Mesinger} A.,  {Dijkstra} M.,  2008, \mnras, 390, 1071

\bibitem[\protect\citeauthoryear{{Mesinger}, {Ewall-Wice} \&
  {Hewitt}}{{Mesinger} et~al.}{2014}]{ME-WH14}
{Mesinger} A.,  {Ewall-Wice} A.,    {Hewitt} J.,  2014, \mnras; arXiv:1310.0465

\bibitem[\protect\citeauthoryear{{Mesinger}, {Ferrara} \& {Spiegel}}{{Mesinger}
  et~al.}{2013}]{MFS13}
{Mesinger} A.,  {Ferrara} A.,    {Spiegel} D.~S.,  2013, \mnras, 431, 621

\bibitem[\protect\citeauthoryear{{Mesinger} \& {Furlanetto}}{{Mesinger} \&
  {Furlanetto}}{2007}]{MF07}
{Mesinger} A.,  {Furlanetto} S.,  2007, \apj, 669, 663

\bibitem[\protect\citeauthoryear{{Mesinger}, {Furlanetto} \& {Cen}}{{Mesinger}
  et~al.}{2011}]{MFC11}
{Mesinger} A.,  {Furlanetto} S.,    {Cen} R.,  2011, \mnras, 411, 955

\bibitem[\protect\citeauthoryear{{Miralda-Escud{\'e}}, {Haehnelt} \&
  {Rees}}{{Miralda-Escud{\'e}} et~al.}{2000}]{MHR00}
{Miralda-Escud{\'e}} J.,  {Haehnelt} M.,    {Rees} M.~J.,  2000, \apj, 530, 1

\bibitem[\protect\citeauthoryear{{Mitra}, {Choudhury} \& {Ferrara}}{{Mitra}
  et~al.}{2015}]{MCF15}
{Mitra} S.,  {Choudhury} T.~R.,    {Ferrara} A.,  2015, ArXiv
  e-prints:1505.05507

\bibitem[\protect\citeauthoryear{{Mondal}, {Bharadwaj}, {Majumdar}, {Bera} \&
  {Acharyya}}{{Mondal} et~al.}{2015}]{Mondal15}
{Mondal} R.,  {Bharadwaj} S.,  {Majumdar} S.,  {Bera} A.,    {Acharyya} A.,
  2015, \mnras, 449, L41

\bibitem[\protect\citeauthoryear{{Morales}}{{Morales}}{2005}]{Morales05}
{Morales} M.~F.,  2005, \apj, 619, 678

\bibitem[\protect\citeauthoryear{{Mutch}, {Geil}, {Poole}, {Angel}, {Duffy},
  {Mesinger} \& {Wyithe}}{{Mutch} et~al.}{2015}]{Mutch15}
{Mutch} S.~J.,  {Geil} P.~M.,  {Poole} G.~B.,  {Angel} P.~W.,  {Duffy} A.~R.,
  {Mesinger} A.,    {Wyithe} J.~S.~B.,  2015, ArXiv e-prints:1512.00562

\bibitem[\protect\citeauthoryear{{Noh} \& {McQuinn}}{{Noh} \&
  {McQuinn}}{2014}]{NM14}
{Noh} Y.,  {McQuinn} M.,  2014, \mnras, 444, 503

\bibitem[\protect\citeauthoryear{{Okamoto}, {Gao} \& {Theuns}}{{Okamoto}
  et~al.}{2008}]{OGT08}
{Okamoto} T.,  {Gao} L.,    {Theuns} T.,  2008, \mnras, 390, 920

\bibitem[\protect\citeauthoryear{{Osterbrock}}{{Osterbrock}}{1989}]{Osterbrock89}
{Osterbrock} D.~E.,  1989, {Astrophysics of gaseous nebulae and active galactic
  nuclei}.
Research supported by the University of California, John Simon Guggenheim
  Memorial Foundation, University of Minnesota, et al.~Mill Valley, CA,
  University Science Books, 1989, 422 p.

\bibitem[\protect\citeauthoryear{{Pacucci}, {Mesinger}, {Mineo} \&
  {Ferrara}}{{Pacucci} et~al.}{2014}]{Pacucci14}
{Pacucci} F.,  {Mesinger} A.,  {Mineo} S.,    {Ferrara} A.,  2014, \mnras, 443,
  678

\bibitem[\protect\citeauthoryear{{Parsons}, {Pober}, {McQuinn}, {Jacobs} \&
  {Aguirre}}{{Parsons} et~al.}{2012}]{Parsons12}
{Parsons} A.,  {Pober} J.,  {McQuinn} M.,  {Jacobs} D.,    {Aguirre} J.,  2012,
  \apj, 753, 81

\bibitem[\protect\citeauthoryear{Parsons et~al.,}{Parsons
  et~al.}{2010}]{Parsons10}
Parsons A.~R.,  et~al., 2010, \aj, 139, 1468

\bibitem[\protect\citeauthoryear{{Parsons} et~al.,}{{Parsons}
  et~al.}{2014}]{Parsons14}
{Parsons} A.~R.,  et~al., 2014, \apj, 788, 106

\bibitem[\protect\citeauthoryear{{Pawlik} \& {Schaye}}{{Pawlik} \&
  {Schaye}}{2009}]{PS09}
{Pawlik} A.~H.,  {Schaye} J.,  2009, \mnras, 396, L46

\bibitem[\protect\citeauthoryear{{Planck Collaboration}, {Ade}, {Aghanim},
  {Arnaud}, {Ashdown}, {Aumont}, {Baccigalupi}, {Banday}, {Barreiro} \& et
  al.}{{Planck Collaboration} et~al.}{2015}]{Planck15}
{Planck Collaboration} {Ade} P.~A.~R.,  {Aghanim} N.,  {Arnaud} M.,  {Ashdown}
  M.,  {Aumont} J.,  {Baccigalupi} C.,  {Banday} A.~J.,  {Barreiro} R.~B.,
  et al. 2015, ArXiv e-prints:1502.01589

\bibitem[\protect\citeauthoryear{Pober et~al.,}{Pober  et~al.}{2013}]{Pober13}
Pober J.~C.,  et~al., 2013, The Astrophysical Journal Letters, 768, L36

\bibitem[\protect\citeauthoryear{{Pober} et~al.,}{{Pober}
  et~al.}{2014}]{Pober14}
{Pober} J.~C.,  et~al., 2014, \apj, 782, 66

\bibitem[\protect\citeauthoryear{{Pritchard} \& {Furlanetto}}{{Pritchard} \&
  {Furlanetto}}{2006}]{PF06}
{Pritchard} J.~R.,  {Furlanetto} S.~R.,  2006, \mnras, 367, 1057

\bibitem[\protect\citeauthoryear{{Pritchard} \& {Furlanetto}}{{Pritchard} \&
  {Furlanetto}}{2007}]{PF07}
{Pritchard} J.~R.,  {Furlanetto} S.~R.,  2007, \mnras, 376, 1680

\bibitem[\protect\citeauthoryear{{Pritchard} \& {Loeb}}{{Pritchard} \&
  {Loeb}}{2012}]{PL12}
{Pritchard} J.~R.,  {Loeb} A.,  2012, Reports on Progress in Physics, 75,
  086901

\bibitem[\protect\citeauthoryear{{Rahmati}, {Pawlik}, {Rai{\v c}evi{\'c}} \&
  {Schaye}}{{Rahmati} et~al.}{2013}]{Rahmati13}
{Rahmati} A.,  {Pawlik} A.~H.,  {Rai{\v c}evi{\'c}} M.,    {Schaye} J.,  2013,
  \mnras, 430, 2427

\bibitem[\protect\citeauthoryear{{Ricotti} \& {Ostriker}}{{Ricotti} \&
  {Ostriker}}{2004}]{RO04}
{Ricotti} M.,  {Ostriker} J.~P.,  2004, \mnras, 352, 547

\bibitem[\protect\citeauthoryear{{Santos}, {Silva}, {Pritchard}, {Cen} \&
  {Cooray}}{{Santos} et~al.}{2011}]{Santos11}
{Santos} M.~G.,  {Silva} M.~B.,  {Pritchard} J.~R.,  {Cen} R.,    {Cooray} A.,
  2011, \aap, 527, A93

\bibitem[\protect\citeauthoryear{{Shapiro}, {Giroux} \& {Babul}}{{Shapiro}
  et~al.}{1994}]{SGB94}
{Shapiro} P.~R.,  {Giroux} M.~L.,    {Babul} A.,  1994, \apj, 427, 25

\bibitem[\protect\citeauthoryear{{Shull} \& {van Steenberg}}{{Shull} \& {van
  Steenberg}}{1985}]{SvS85}
{Shull} J.~M.,  {van Steenberg} M.~E.,  1985, \apj, 298, 268

\bibitem[\protect\citeauthoryear{{Smith} et~al.,}{{Smith}
  et~al.}{2016}]{Smith16}
{Smith} B.~M.,  et~al., 2016, ArXiv e-prints:1602.01555

\bibitem[\protect\citeauthoryear{{Sobacchi} \& {Mesinger}}{{Sobacchi} \&
  {Mesinger}}{2013}]{SM13a}
{Sobacchi} E.,  {Mesinger} A.,  2013, \mnras, 432, L51

\bibitem[\protect\citeauthoryear{{Sobacchi} \& {Mesinger}}{{Sobacchi} \&
  {Mesinger}}{2014}]{SM14}
{Sobacchi} E.,  {Mesinger} A.,  2014, \mnras, 440, 1662

\bibitem[\protect\citeauthoryear{{Springel} \& {Hernquist}}{{Springel} \&
  {Hernquist}}{2003}]{SH03}
{Springel} V.,  {Hernquist} L.,  2003, \mnras, 339, 312

\bibitem[\protect\citeauthoryear{{Sun} \& {Furlanetto}}{{Sun} \&
  {Furlanetto}}{2015}]{SF15}
{Sun} G.,  {Furlanetto} S.~R.,  2015, ArXiv e-prints:1512.06219

\bibitem[\protect\citeauthoryear{{Thompson}, {Moran} \& {Swenson}}{{Thompson}
  et~al.}{2007}]{TMS07}
{Thompson} A.~R.,  {Moran} J.~M.,    {Swenson} G.~W.,  2007, {Interferometry
  and Synthesis in Radio Astronomy, John Wiley \& Sons, 2007.}

\bibitem[\protect\citeauthoryear{{Thoul} \& {Weinberg}}{{Thoul} \&
  {Weinberg}}{1996}]{TW96}
{Thoul} A.~A.,  {Weinberg} D.~H.,  1996, \apj, 465, 608

\bibitem[\protect\citeauthoryear{{Tingay} et~al.,}{{Tingay}
  et~al.}{2013}]{Tingay13}
{Tingay} S.~J.,  et~al., 2013, \pasa, 30, 7

\bibitem[\protect\citeauthoryear{{Vald{\'e}s}, {Evoli} \&
  {Ferrara}}{{Vald{\'e}s} et~al.}{2010}]{VEF11}
{Vald{\'e}s} M.,  {Evoli} C.,    {Ferrara} A.,  2010, \mnras, 404, 1569

\bibitem[\protect\citeauthoryear{{van Haarlem} et~al.,}{{van Haarlem}
  et~al.}{2013}]{vanHaarlem13}
{van Haarlem} M.~P.,  et~al., 2013, \aap, 556, A2

\bibitem[\protect\citeauthoryear{{Wouthuysen}}{{Wouthuysen}}{1952}]{Wouthuysen52}
{Wouthuysen} S.~A.,  1952, \aj, 57, 31

\bibitem[\protect\citeauthoryear{{Zel'Dovich}}{{Zel'Dovich}}{1970}]{Zeldovich70}
{Zel'Dovich} Y.~B.,  1970, \aap, 5, 84

\end{thebibliography}

\end{document}